\documentclass[aps,prx,showpacs,floatfix,twocolumn,superscriptaddress]{revtex4-2}

\usepackage{float}
\usepackage{color}
\usepackage{bm}
\usepackage{hyperref}
\usepackage{todonotes}
\usepackage{verbatim}
\usepackage{soul}
\usepackage[version=4]{mhchem}
\usepackage[no-test-for-array]{nicematrix}

\usepackage{glossaries}
\usepackage{sidecap}
\usepackage{hyperref}
\usepackage{verbatim}
\hypersetup{
    colorlinks=true,
    linkcolor=blue,
    filecolor=magenta,      
    urlcolor=red,
    citecolor=blue,
}

\def\Q{\ensuremath{\bm{Q}}}
\newcommand{\angstrom}{\mbox{\normalfont\AA}}

\newacronym{RIXS}{RIXS}{resonant inelastic x-ray scattering}
\newacronym{FWHM}{FWHM}{full-width at half-maximum}
\newacronym{HWHM}{HWHM}{half-width at half-maximum}
\newacronym{AIM}{AIM}{Anderson impurity model}
\newacronym{ED}{ED}{exact diagonalization}
\newacronym{FM}{FM}{ferromagnetic}
\newacronym{DFT}{DFT}{density functional theory}
\newacronym{2D}{2D}{two-dimensional}
\newacronym{vdW}{vdW}{van der Waals}
\newacronym{XAS}{XAS}{x-ray absorption spectroscopy}

\begin{document}

\title{Dispersive dark excitons in van der Waals ferromagnet CrI$_3$}

\author{W. He}\email{Current address: Stanford Institute for Materials and Energy Sciences, SLAC National Accelerator Laboratory, Menlo Park, CA 94025, USA; weihe@stanford.edu}
\author{J. Sears}
\affiliation{Department of Condensed Matter Physics and Materials Science, Brookhaven National Laboratory, Upton, New York 11973, USA}

\author{F. Barantani}
\affiliation{Department of Physics, The University of Texas at Austin, Austin, Texas, USA, 78712}

\author{T. Kim}
\affiliation{National Synchrotron Light Source II, Brookhaven National Laboratory, Upton, New York 11973, USA}

\author{J. W. Villanova}
\author{T. Berlijn}
\affiliation{Center for Nanophase Materials Sciences, Oak Ridge National Laboratory, Oak Ridge, Tennessee 37831, USA}

\author{M. Lajer}
\affiliation{Department of Condensed Matter Physics and Materials Science, Brookhaven National Laboratory, Upton, New York 11973, USA}

\author{M. A. McGuire}
\affiliation{Materials Science and Technology Division, Oak Ridge National Laboratory, 1 Bethel Valley Road, Oak Ridge, Tennessee 37831, USA}

\author{J. Pelliciari}
\author{V. Bisogni}
\affiliation{National Synchrotron Light Source II, Brookhaven National Laboratory, Upton, New York 11973, USA}

\author{S.~Johnston}
\affiliation{Department of Physics and Astronomy, The University of Tennessee, Knoxville, Tennessee 37966, USA\looseness=-1}
\affiliation{Institute for Advanced Materials and Manufacturing, The University of Tennessee, Knoxville, Tennessee 37996, USA\looseness=-1}

\author{E. Baldini}
\affiliation{Department of Physics, The University of Texas at Austin, Austin, Texas, USA, 78712}

\author{M. Mitrano}
\affiliation{Department of Physics, Harvard University, Cambridge, Massachusetts 02138, USA\looseness=-1}

\author{M. P. M. Dean}
\email{mdean@bnl.gov}
\affiliation{Department of Condensed Matter Physics and Materials Science, Brookhaven National Laboratory, Upton, New York 11973, USA}

\date{\today}

\begin{abstract}

Spin-flip dark excitons are optical-dipole-forbidden quasiparticles with remarkable potential in optoelectronics, especially when they are realized within cleavable van der Waals materials. Despite this potential, dark excitons have not yet been definitively identified in ferromagnetic van der Waals materials. Here, we report two dark excitons in a model ferromagnetic material \ce{CrI3} using high-resolution resonant inelastic x-ray scattering (RIXS) and show that they feature narrower linewidths compared to the bright excitons previously reported in this material. These excitons are shown to have spin-flip character, to disperse as a function of momentum, and to change through the ferromagnetic transition temperature. Given the versatility of van der Waals materials, these excitons hold promise for new types of magneto-optical functionality.

\end{abstract}

\maketitle

\section{Introduction}

Excitons play a key role in determining the optical properties of solids, and their strong light-matter coupling paves the way for exploring new aspects of many-body physics \cite{Koch2006semiconductor}. Dark excitons are particularly interesting because they involve optical-dipole-forbidden transitions \cite{Kitzmann2022Spinflip}. For this reason, they have reduced rates of radiative recombination and enhanced lifetimes and in many cases they can be sensitively controlled by external means, such as magnetic field \cite{Zhang2017Magnetic}. These properties endow them with great potential in quantum information storage and communication \cite{Poem2010Accessing}.

Early studies of dark excitons began in the 1990s with quantum dots \cite{Nirmal1995observation}, followed by research on organic materials \cite{Congreve2013external}, and later expanded to transition-metal dichalcogenides \cite{Wang2018Colloquium}. The recent discovery of magnetic \gls*{vdW} materials provides a new platform for studying excitons \cite{Burch2018Review, Gong2019Two,Wang2022Review,Ahn2024Progress,Mak2019probing} and fascinating interactions between magnetism and excitons have been observed in several antiferromagnetic systems \cite{Kang2020Coherent, Bae2022exciton, He2024Magnetically, DiScala2024Elucidating, Occhialini2024Nature}. Understanding the electronic structure of these excitons and their interactions with magnetism is not only interesting from a fundamental point of view, but it may also offer new types of magneto-optical functionality such as optical read-out of magnetic states or quantum sensors \cite{Wilson2021excitons}. \ce{CrI3} provides an opportunity to study excitons in a \gls*{FM} \gls*{vdW} material even down to the monolayer limit \cite{Huang2017Layer, Jin2020Observation}. Optical studies have revealed several bright excitons around $1.50$, $1.85$ and $2.2$ eV in this material alongside several other optical features \cite{Pollini1970intrinsic, Bermudez1979spectroscopic, Seyler2018Ligandfield, Jin2020Observation}; however dark excitons have not been definitively identified. 

\Gls*{RIXS} is directly sensitive to optically forbidden excitations and has recently emerged as a powerful probe of excitons and their interactions with magnetism, in several magnetic \gls*{vdW} materials \cite{Kang2020Coherent, He2024Magnetically, DiScala2024Elucidating, Son2022multiferroic, Occhialini2024Nature, Mitrano2024exploring}. 
In this Letter, we use Cr $L_3$-edge \gls*{RIXS} to identify two dark excitons near $1.7$~eV in \ce{CrI3}. Both dark excitons are much sharper than other bright excitons and disperse with a bandwidth ($\sim 10$ meV) similar to the energy scale of magnetic exchange interactions in this material. Together with the change of their intensities across the \gls*{FM} ordering temperature $T_{\mathrm c}$, our experimental findings suggest an intimate relationship between \ce{CrI3}'s dark excitons and its magnetism. The electronic character of the dark excitons is borne out by our \gls*{ED} calculations, which reveal that these excitons are spin-flip transitions in nature and are predominantly governed by Hund's coupling.

\section{Methods}

Bulk single crystals of \ce{CrI3} were synthesized by reacting the elements together in an evacuated fused silica ampoule \cite{McGuire2015Coupling}. \ce{CrI3} undergoes a structural phase transition between the high-temperature monoclinic structure (space group $C2/m$, \#12) and the low-temperature rhomobohedral structure (space group $R\bar{3}$, \#148) over a broad temperature range ($100$--$220$~K) upon thermal cycling. The temperature of the sample was kept at $T = 30$~K, deep into the \gls*{FM} phase of the material, unless otherwise specified. We consequently used the rhomobohedral unit cell notation with lattice parameters $a = b = 6.867$~\angstrom{}, $c = 19.807$~\angstrom{}, and $\gamma =  120^\circ$ throughout the manuscript, and index reciprocal space in terms of scattering vector $\Q{}=(H,K,L)$ in reciprocal lattice units (r.l.u.). 

To avoid sample degradation in air, we mounted the sample on a copper sample holder in a glove box, cleaved with scotch tape in \ce{N2} atmosphere ($\sim 3$\% relative humidity level) to expose a fresh surface, and directly transferred the cleaved sample into the \gls*{RIXS} sample chamber. The surface normal of the sample was parallel to the $c$-axis. The in-plane orientation was determined by checking the residue on the scotch tape with a laboratory single-crystal x-ray diffractometer.

Cr $L_3$-edge \gls*{RIXS} measurements were performed at the SIX 2-ID beamline of the National Synchrotron Light Source II. Data were taken with linear horizontal ($\pi$) polarization in the $(H0L)$ scattering plane unless otherwise specified. The spectrometer was operated with a high energy resolution of $30.5$~meV \gls*{FWHM} (the exit slit size was $30$~{\textmu}m). Since the interlayer coupling in \ce{CrI3} is weak, we fixed the scattering angle at $2\Theta=150^\circ$ and expressed \Q{} in terms of the projected in-plane component of the momentum $H$. An angle-dependent self-absorption correction \cite{Miao2017High} was applied to the \gls*{RIXS} spectra, which, however, does not affect exciton energies or relative intensity changes at fixed \Q{}. The \gls*{XAS} spectra were taken using the partial fluorescence yield mode with the \gls*{RIXS} detector, which covers the energy loss up to $\sim 11$~eV. The exit slit size was much larger ($300$~{\textmu}m) for the \gls*{XAS} measurements to increase the flux.

\section{Identification of dark excitons}

\begin{figure}
\includegraphics{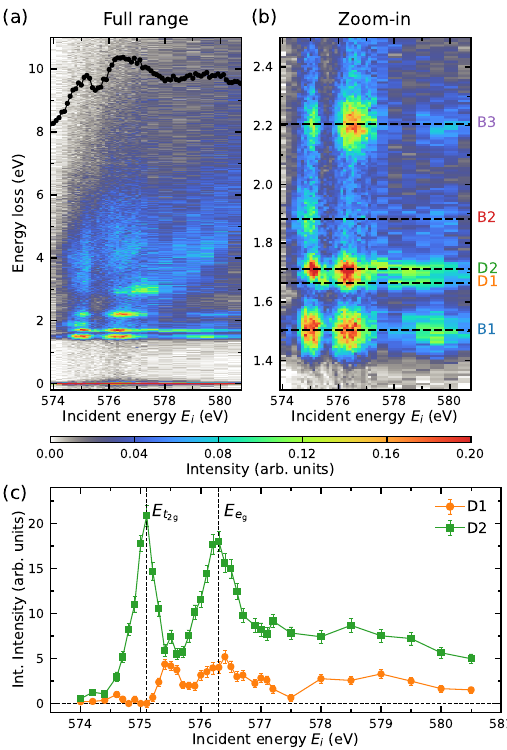}
\caption{Resonance behavior of dark excitons. (a) Cr $L_3$-edge \gls*{RIXS} incident energy map taken at $T=30$~K with $\pi$-polarized x-rays incident on the sample at $\theta = 14.5^\circ$ and scattered to $2\Theta = 150^\circ$ in the ($H0L$) scattering plane, corresponding to $H=-0.46$~r.l.u. The overlaid black curve on the top is \gls*{XAS} spectrum taken at the same conditions (including x-ray polarization, experimental geometry, and temperature). (b) Zoom of the exciton resonances. Horizontal dashed lines are fitted exciton energies. Two peaks near $1.7$~eV are identified as the dark excitons and denoted as D1 and D2. The other three peaks are bright excitons previously observed in optical measurements \cite{Pollini1970intrinsic, Bermudez1979spectroscopic, Seyler2018Ligandfield, Jin2020Observation} and therefore denoted as B1--B3. (c) The fitted integrated intensities of the two dark excitons as a function of incident photon energy $E_i$ through the $E_{t_{\mathrm{2g}}}$ and $E_{e_{\mathrm{g}}}$ resonances. D1 resonates at $E_{e_{\mathrm{g}}}$, whereas D2 resonates at $E_{e_{\mathrm{g}}}$ and $E_{t_{\mathrm{2g}}}$. As shown in Supplemental Material Sec.~S1, these effects show minimal dichroism \cite{supp}. 
}
\label{fig:energy_map}
\end{figure}

\begin{figure*}
\includegraphics{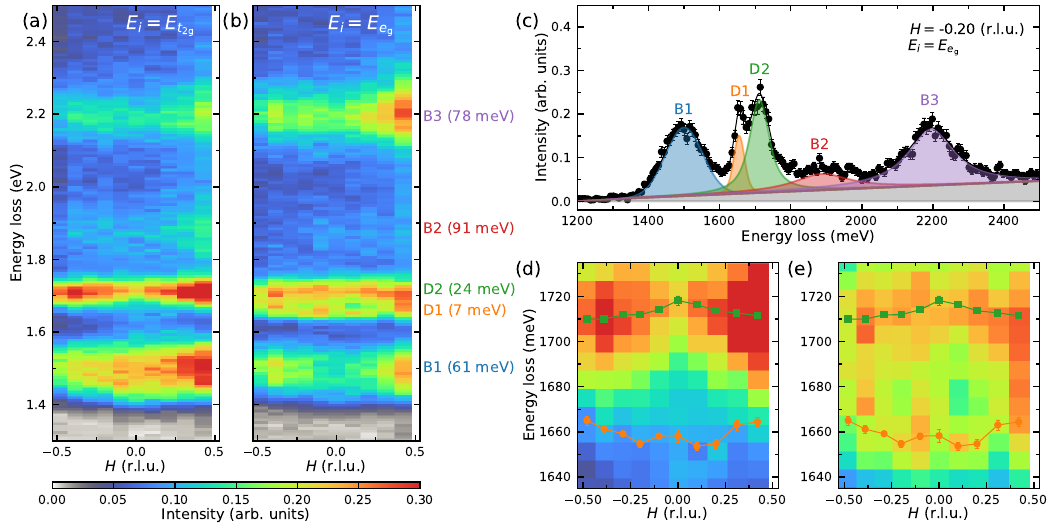}
\caption{Dispersion of the dark excitons. (a),(b) \gls*{RIXS} intensity map at $T=30$~K as a function of in-plane momentum transfer $H$ measured with (a) $E_i = E_{t_{\mathrm{2g}}}$ where the D2 exciton is strongest and (b) $E_i = E_{e_{\mathrm{g}}}$ where the D1 and D2 excitons are visible. The two dark excitons are much narrower than the other bright excitons, as seen by inspecting the intrinsic resolution deconvolved \gls*{HWHM} of the peaks as included in brackets after the peak labels. (c) Representative fit at $H=-0.20$ r.l.u.\ with $E_i = E_{e_{\mathrm{g}}}$. (d),(e) Zoom of the dark exciton dispersion at (d) $E_i = E_{t_{\mathrm{2g}}}$ and (e) $E_i = E_{e_{\mathrm{g}}}$. For each momentum, we co-fit the spectra taken at the two resonances with shared exciton energies and widths. The co-fitted energy dispersion overlays the colormaps.
}
\label{fig:dispersion}
\end{figure*}

Figure~\ref{fig:energy_map}(a) shows Cr $L_3$-edge \gls*{RIXS} spectra of \ce{CrI3} as a function of incident x-ray energy. Peaks below $2.5$~eV energy loss are mainly local transitions within the Cr $3d$ orbital manifold, while broad features at higher energy can be ascribed to charge transfer processes that heavily involve ligand orbitals, and x-ray fluorescence arising from more extended states \cite{Ament2011Resonant}. In the 1--3 eV energy window depicted in Fig.~\ref{fig:energy_map}(b), three peaks, located at $1.50$~eV, $1.88$~eV, and $2.21$~eV, match the exciton energies previously reported in optical experiments \cite{Pollini1970intrinsic, Bermudez1979spectroscopic, Seyler2018Ligandfield, Jin2020Observation}, so we denote them as bright excitons B1--B3. More excitingly, two additional features are observed at $1.66$~eV and $1.71$~eV, which were not seen in prior \gls*{RIXS} measurements due to their lower resolution ($180$/$350$~meV ared to $30.5$~meV used here \cite{Ghosh2023Magnetic, Shao2021spectroscopic}). These modes have not yet been definitively identified in optical spectra, so, following standard terminology in optics, we will refer to these as dark excitons and denote them as D1 and D2. We note that there are no additional features at energies below B1, contrary to an earlier prediction that the lowest energy dark excitons should exist around $0.9$~eV \cite{Wu2019Physical}. 

Figure~\ref{fig:energy_map}(b) exhibits two prominent resonances at $E_i = 575.1$ eV and $576.3$ eV. Although there is strong $t_{\mathrm{2g}}$-$e_{\mathrm{g}}$ mixing in \ce{CrI3}, these two resonances correspond to more $t_{2g}$-like and more $e_g$-like orbital manifolds, respectively, as shown in Supplemental Fig.~S16. We therefore label them as $E_{t_{\mathrm{2g}}}$ and $E_{e_{\mathrm{g}}}$ resonances hereafter. As shown in Fig.~\ref{fig:energy_map}(c), D2 resonates with $t_{\mathrm{2g}}$ and $e_{\mathrm{g}}$ intermediate states, but D1 only resonates at the $e_{\mathrm{g}}$ condition. An additional small resonance is present around 575.5~eV, which is generated by the exchange part of the core-valence Coulomb interaction on the Cr site. We will use the excitons' energy and angular dependence to identify their electronic character later in this Article.

\section{Dark excitons dispersion}
\begin{figure}
\includegraphics{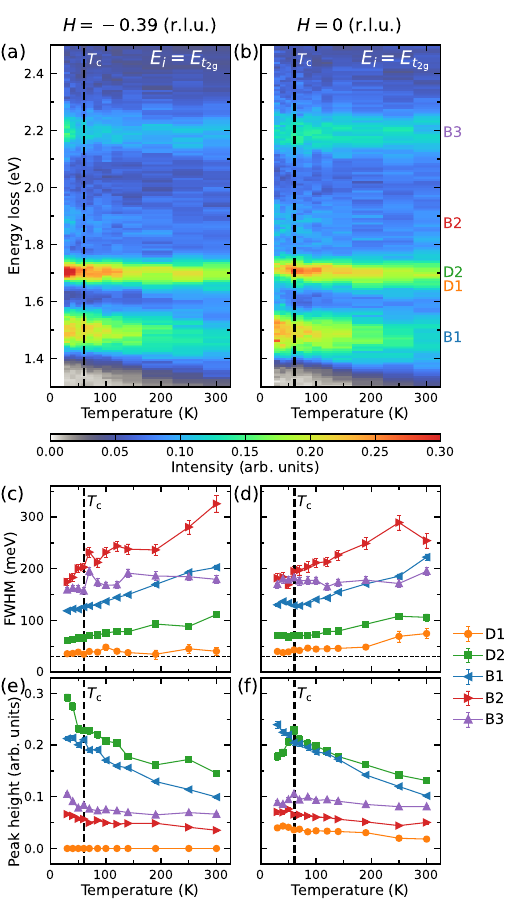}
\caption{Temperature dependence of the dark excitons. (a),(b) \gls*{RIXS} intensity map as a function of temperature measured at two different in-plane momenta ($H=-0.39$ and $H=0$ r.l.u., respectively) with the same incident energy $E_{t_{\mathrm{2g}}}$. (c)--(f) Corresponding \gls*{FWHM} and peak height extracted from the fits as a function of temperature. Changes in peak height through $T_c$ primarily reflect changes in the integrated intensity of the peaks. We plot the peak intensity here because this quantity can be determined more precisely. Error bars represent one standard deviation. The fits were performed by co-fitting spectra taken at two incident energies with shared exciton energies and widths for each momentum and temperature. The data taken at the other incident energy $E_{e_{\mathrm{g}}}$ are shown in Supplemental Material Sec.~S2 \cite{supp}. The vertical black lines indicate the \gls*{FM} transition temperature $T_{\mathrm c}$ and the horizontal black lines in (c)\&(d) indicate the energy resolution.
}
\label{fig:Tdep}
\end{figure}
Having identified the existence of dark excitons in \ce{CrI3}, we explore their propagation by mapping out their in-plane dispersion at the two resonant energies $E_{t_{\mathrm{2g}}}$ and $E_{e_{\mathrm{g}}}$, as presented in Fig.~\ref{fig:dispersion}(a) and (b). Intriguingly, both excitons D1 and D2 exhibit a small dispersion but with opposite trends. We co-fit the spectra at the two different resonances, as described in Supplemental Material Sec.~S3, and shown in Fig.~\ref{fig:dispersion}(c). The fitted exciton energies in Fig.~\ref{fig:dispersion}(d)\&(e) confirm the presence of dispersion with similar bandwidths of $\sim 10$~meV. Such bandwidths are comparable to the energy scale of the magnon dispersion \cite{Chen2018Topological}, hinting at the involvement of exciton-magnon interactions when these dark excitons propagate in the lattice. The widths of the two dark excitons are dramatically sharper than those of other bright excitons, especially D1, which is almost resolution limited and about ten times narrower than the bright excitons [see Fig.~\ref{fig:dispersion}(b)]. The three bright excitons do not exhibit any detectable dispersion (see Supplemental Fig.~S9), likely due to their broader linewidths.

\begin{figure*}
\includegraphics{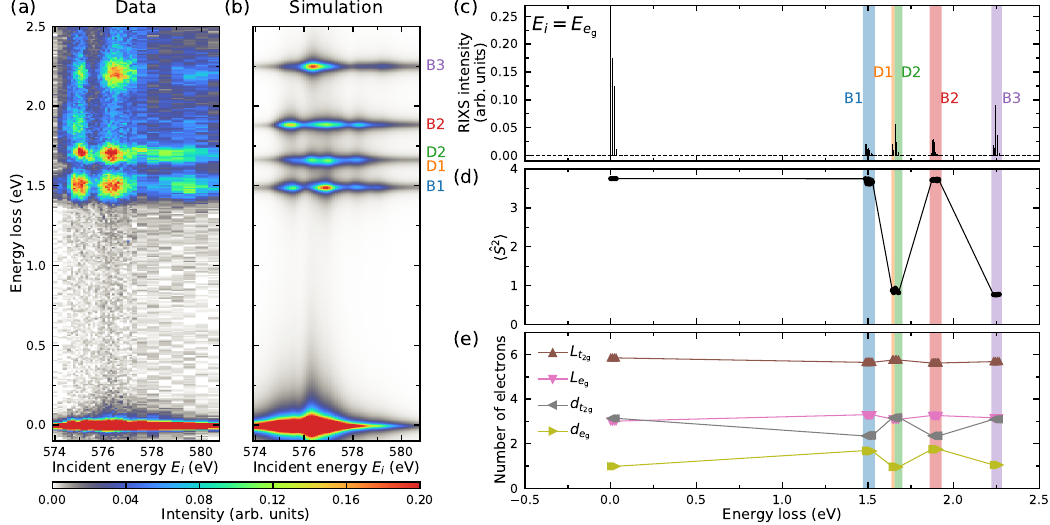}
\caption{Electronic character of the dark excitons. (a) \gls*{RIXS} intensity map as a function of incident photon energy through the Cr $L_3$ resonance. This is the same data presented in Fig.~\ref{fig:energy_map} with an energy window chosen to highlight the low-energy excitations. (b) \gls*{RIXS} calculations that reproduce the energy and resonant profile of the five lowest energy excitons in the material. (c) Calculated \gls*{RIXS} stick diagram at the main resonant energy $E_{e_{\mathrm{g}}}$. (d),(e) Analysis of the ground and excited states for the calculated \gls*{RIXS} spectrum. (d) Expectation value of the total spin operator squared $\langle \hat{S}^2 \rangle$. (e) Electron occupations of Cr $3d$ (denoted by $d$) and ligand (denoted by $L$) orbitals. 
}
\label{fig:ED}
\end{figure*}

\section{Temperature dependence}

Next, we measured the temperature dependence of \gls*{RIXS} spectra at $E_i = E_{t_{\mathrm{2g}}}$ through the \gls*{FM} transition temperature $T_{\mathrm c}=61$~K up to room temperature. Figure~\ref{fig:Tdep} plots data and fits at two momenta. All excitons show an overall trend toward larger widths at higher temperatures. Such behavior is similar to \gls*{RIXS} measurements of excitons in other related materials such as \ce{NiPS3} \cite{Kang2020Coherent} and nickel dihalides \cite{Son2022multiferroic, Occhialini2024Nature}. D2 (and to a lesser extent B3) shows a clear anomaly around the \gls*{FM} transition temperature $T_{\mathrm c}$. The trends are even opposite through $T_{\mathrm c}$ at the two selected momenta, demonstrating an unusual momentum-dependent coupling between magnetism and dark exciton states. Other bright excitons, such as B1, also exhibit anomalies across $T_{\mathrm c}$ at certain momenta and incident energies (See Supplemental Fig.~S3). The momentum dependence of these anomalies indicates underlying physics that cannot be fully captured by established cluster-based methods commonly used to interpret \gls*{RIXS}. We also note that B1 (and to a lesser extent D2) softens steadily upon warming, particular above $T_{\mathrm c}$, which is reminiscence of the exciton behavior in \ce{NiPS3} and \ce{NiI2} \cite{Kang2020Coherent, Son2022multiferroic}. Such interesting behavior might be related to the electron-phonon interactions, which was used to explain the electronic gap shift observed in optics \cite{Tomarchio2021Low}.

\section{Electronic character of excitons}
An advantage of \gls*{RIXS} compared to optics is that it couples directly to dipole-forbidden transitions in a well-known way, allowing us to extract the electronic character of the excitons. Indeed, ambiguities in the optical cross section for these excitons has led to different suggestions for the types of excitations present in chromium trihalides in 1.50--1.85~eV window including features coming from trigonal crystal fields to vibronic structures \cite{Pollini1970intrinsic, Bermudez1979spectroscopic}. To interpret the present spectra, we built an \gls*{AIM} for \ce{CrI3} and computed the \gls*{RIXS} spectrum using \gls*{ED} methods \cite{Wang2019EDRIXS}. The model includes Coulomb repulsion, Hund's coupling, crystal field, and hopping terms derived from the Wannierization of our \ce{CrI3} \gls*{DFT} calculations as detailed in Appendix~\ref{sec:ED}. The spectra, shown in Fig.~\ref{fig:ED}(a) and (b), capture the exciton energies including the double-peak structure of the two dark excitons and the overall trends in the resonances.

To identify the excitons, we report the final state expectation values of the spin-squared $\langle \hat{S}^2 \rangle$ and electron population operators in Fig.~\ref{fig:ED}(c)-(e). We note that it is vital to have a small charge-transfer energy (in fact, the best fit is achieved using $\Delta=-1.3$~eV with an error bar of $\sim 1$~eV) to obtain the correct energies for the entire spectrum. Such a small (or even slightly negative) charge transfer energy leads to approximately four electrons in the $d$ states and approximately one hole occupying the ligand $e_g$ orbitals not only in the ground states but also in the low energy excitations, as shown in Fig.~\ref{fig:ED}(e), and in accordance with previous \gls*{DFT}-based results \cite{Wu2019Physical, Kashin2020Orbitally, Acharya2021Electronic, Kvashnin2022Dynamical}. The importance of charge-transfer is further underlined by the fact that a single-site atomic model cannot adequately fit the spectrum (see Supplemental Material Sec.~5A \cite{supp}). As such, the excitations observed here have substantial I character and are not strictly $dd$-excitations. The non-dispersive modes observed here are well-described by the broader concept of ligand-field excitons \cite{Ballhausen1962introduction}. However, since these are by definition local, the dispersive D1 and D2 are, in our opinion, best termed as excitons, although all these terms share some similarities \footnote{See the supplemental Material Section S6 for more discussion of this issue.}.

The ground state features half-filled $t_{\mathrm{2g}}$ orbitals, with $\langle \hat{S}^2 \rangle \approx3.75$, corresponding to a high-spin configuration with $S=3/2$. The D1 and D2 dark excitons involve only a small change in electron population, with the largest component of charge motion being about 0.1 electrons moving from the ligand $L_{t_{2g}}$ to the $L_{e_g}$ states, the change from $d$ to $L$ states is even smaller at 0.03 electrons. More notable, this transition involves a low-spin configuration final state close to $S=1/2$. Consequently, the energy scale of these dark excitons is determined mainly by the Hund's coupling for Cr $3d$ orbitals, consistent with their spin-flip character. Although D1 and D2 have dominant $t_{2g}$ character, they feature non-zero mixing with the $e_g$ manifold [see Fig.~\ref{fig:ED}(e)]. This mixing endows D1 and D2 with a distinct orbital angular momentum, leading to their small energetic splitting and their distinct resonant profile at the $t_{2g}$ and $e_g$ conditions. The nearby bright excitons B1 and B2 are both crystal field transitions, conserving spin and moving an electron from the $t_{2g}$ to the $e_g$ states. Their energies are determined primarily by a combination of crystal field $10D_q$ and charge-transfer energy $\Delta$ (see Appendix~\ref{sec:ED} for the definitions of these quantities).

The higher-energy B3 exciton is again a spin-flip transition but with different symmetry that also redistributes a small amount of weight from the $t_{2g}$ to $e_g$ states. Its intensity exhibits a temperature dependence similar to that of the dark exciton D2 (although the change is less dramatic) in Fig.~\ref{fig:Tdep}(e) and (f), possibly implying their similar spin-flip character. B3 is particularly broad, suggesting that this mode may be coupling to other excitations. Such a coupling could play a role in the finite optical cross-section for this mode, which is larger than expected since it is a spin-flip transition. It is also possible that extended lattice models could be required to understand the detailed nature of the B2 and B3 excitons and their optical-cross section more fully, as suggested by many-body perturbation-theory plus Bethe-Salpeter equation calculations  \cite{Wu2019Physical, Acharya2022real}.

\section{Discussion and conclusions}
The results here provide the first example of dispersive excitons in a \gls*{FM} \gls*{vdW} material, extending the recent identification of this phenomenon in antiferromagnetic \gls*{vdW} materials such as \ce{NiPS3} \cite{He2024Magnetically} and nickel dihalides \cite{Occhialini2024Nature}. These dispersive excitons are all sharp spin-flip excitations bound by Hund's exchange interactions. The sharpness of the exciton (i.e., its long lifetime) may be related to its spin-flip nature because such transitions do not involve charge redistribution and therefore have minimal couplings to phonons and reduced efficiency of radiative recombination \cite{Jin2020Observation}. Prior works have made similar arguments for \ce{NiPS3} and \ce{NiI2}, where the spin-flip character of excitons is regarded as the major aspect of their physics \cite{Kang2020Coherent, He2024Magnetically, Hamad2024singlet}. The small electron transfer involved in the transition has been suggested to play a minor role, even though the magnitude of electron transfer in \ce{NiPS3} is of order 0.2 electron and about seven times larger than the charge transfer observed here \cite{He2024Magnetically}. 

All these excitons also have appreciable ligand-hole involvement due to the small charge-transfer energy, which may facilitate the exciton propagation in the magnetic background \cite{Occhialini2024Nature}. However, there are also key distinctions between \ce{CrI3} excitons and their counterparts in antiferromagnets. One clear difference is the exciton bandwidth. Phenomenologically, unlike the case in \ce{NiPS3} or nickel dihalides where the exciton bandwidth is significantly smaller than the magnon bandwidth \cite{He2024Magnetically, Occhialini2024Nature}, both dark excitons in \ce{CrI3} have bandwidths comparable to the magnon bands \cite{Chen2018Topological}. In addition, these excitons exhibit a peculiar temperature dependence clearly correlated with $T_{\mathrm c}$. Since \ce{CrI3} exhibits layer-dependent magnetism \cite{Huang2017Layer}, it would be of high interest to examine the exciton behaviors in the few-layer limit of \ce{CrI3} in the near future. We also note that the energies of the \ce{CrI3} dark excitons are higher than the electronic band gap, which was reported to be $1.1$--$1.3$~eV \cite{Tomarchio2021Low} in \ce{CrI3}. This above-band-gap character may explain relatively large linewidth and their darkness in optics due to the strong sloping background coming from interband transitions. As a comparison, the previously discovered spin-flip excitons in \ce{NiPS3} involve below-band-gap transitions \cite{Kim2018Charge,Kang2020Coherent,He2024Magnetically}.

In conclusion, we report two dark excitons in \ce{CrI3} directly measured with high resolution \gls*{RIXS}. Our results showcase \gls*{RIXS} as a powerful tool in studying these dark states with a readily interpretable cross section and large momentum space coverage, complementary to optical measurements. The excitons feature strong coupling with the magnetism---they disperse with bandwidths similar to magnons and display unusual temperature dependence across the magnetic transition temperature. Our results will guide future optical experiments to detect and manipulate these dark excitons in \ce{CrI3}. In the past, various methods have been employed to brighten dark excitons in optical measurements, such as the application of an in-plane magnetic field \cite{Zhang2017Magnetic}, near-field coupling to surface plasmon polaritons \cite{Zhou2017Probing}, or by nano-optical tip-enhanced approaches \cite{Park2018Radiative}. With the better-targeted energies provided by our study, there is a high likelihood of probing these dark excitons in \ce{CrI3} using these advanced optical techniques. Indeed, the possibility of accessing D1 and/or D2 optically is supported by a subset of the prior optical studies, which identified shoulder features in the spectra indicative of modes in the 1.7~eV range \cite{Jin2020Observation}. Moreover, our \gls*{ED} calculations will also inform future theory in more accurately describing the electronic properties of \ce{CrI3}. We believe that our discovery of dark excitons in \ce{CrI3} is just the tip of the iceberg, and the utilization of \gls*{RIXS} will expedite the expansion of this family of materials and foster both the understanding of the fundamental physics and the potential applications of dark excitons in devices.

The supporting data for the plots in this article are openly available from the Zenodo database \cite{repo}.

\begin{acknowledgments}
Work performed at Brookhaven National Laboratory and Harvard University was supported by the U.S.\ Department of Energy (DOE), Division of Materials Science, under Contract No.~DE-SC0012704. Work performed at the University of Texas at Austin was supported by the United States Army Research Office (W911NF-23-1-0394) (F.B.) and the National Science Foundation under the NSF CAREER award 2441874 (E.B.). F.B. acknowledges additional support from the Swiss NSF under fellowship P500PT$\_$214437. S.~J. was supported by the U.S.\ Department of Energy, Office of Science, Office of Basic Energy Sciences, under Award Number DE-SC0022311. Part of this research (T.B.) was conducted at the Center for Nanophase Materials Sciences, which is a DOE Office of Science User Facility. The work by J.W.V.\ is supported by the Quantum Science Center (QSC), a National Quantum Information Science Research Center of DOE. Crystal growth at ORNL as supported by the U.S.\ DOE, Office of Science, Basic Energy Sciences, Material Science and Engineering Division. This research used beamline 2-ID of the National Synchrotron Light Source II, a U.S.\ DOE Office of Science User Facility operated for the DOE Office of Science by Brookhaven National Laboratory under Contract No.~DE-SC0012704. We also acknowledge glovebox resources made available through BNL/LDRD \#19-013.
\end{acknowledgments}

\appendix

\section{\gls*{ED} \gls*{RIXS} calculations\label{sec:ED}}
The \gls*{RIXS} spectra in this work were simulated based on standard \gls*{ED} methods implemented in the open source software EDRIXS \cite{Wang2019EDRIXS}. The \gls*{RIXS} cross section was calculated using the Kramers-Heisenberg formula with the polarization-dependent dipole approximation. The model we employ here is an \gls*{AIM}, which was constructed using the bonding ligand orbitals of a \ce{CrI6} cluster model. Here, we provide details on the parameters that we used and the methods that we employed to determine them. 

In an \gls*{AIM}, we can represent a cluster with fewer orbitals by using symmetry-adapted linear combinations of ligand orbitals \cite{Haverkort2012multiplet}, making the calculation numerically much more efficient with essentially no loss of accuracy. In our case, we have 10 Cr $3d$ spin-orbitals and 10 ligand spin-orbitals with the same symmetry, so there are 13 electrons in total occupying 20 spin-resolved orbitals in the initial and final states. In the intermediate states, the Cr $2p$ orbitals are included to simulate the core hole created in the \gls*{RIXS} process. The calculations were performed in the full basis using the Fortran \gls*{ED} solver provided in EDRIXS \cite{Wang2019EDRIXS}. For the \gls*{RIXS} cross section calculations, we have the experimental geometry explicitly considered, i.e., the scattering angle $2\Theta$ is fixed to $150^\circ$ and the sample angle $\theta$ is kept at $14.5^\circ$. An inverse core-hole lifetime $\Gamma_c = 0.3$~eV \gls*{HWHM} was used to fit the observed width of the resonance and the final state energy loss spectra are broadened using a Lorentzian function with a \gls*{FWHM} of $0.03$~eV, in order to match the observed minimum width of the dark excitons.

The Hamiltonian of the model includes the Coulomb interactions, on-site energy for each orbital, hoppings between different orbitals, spin-orbit coupling, and the Zeeman interaction.

We parameterize the Coulomb interactions via Slater integrals, which include $F^0_{dd}$, $F^2_{dd}$, and $F^4_{dd}$ for the Cr $3d$ orbitals, and $F^0_{dp}$, $F^2_{dp}$, $G^1_{dp}$, and $G^3_{dp}$ for the interactions between Cr $3d$ and $2p$ core orbitals in the intermediate states. $F^0_{dd}$ and $F^0_{dp}$ are related to the Cr $3d$ onsite Coulomb repulsion $U_{dd}$ and core-hole potential $U_{dp}$, respectively, which are discussed later. The rest of the parameters are obtained by starting from their Hartree-Fock atomic values and scaling them down to account for the screening effect in the solids. For simplicity, we used two overall scaling factors, one for Cr $3d$ orbitals ($k_{dd}$) and the other for the core-hole interactions ($k_{dp}$).

Hopping integrals describe hybridization between different orbitals. This can be expressed using a $10 \times 10$ matrix $H_\text{hopping}$. We determined the off-diagonal inter-orbital hopping parameters from first-principles by calculating the electronic structure with the \textsc{VASP} \gls*{DFT} code \cite{Kresse1996effcient, Kresse1996effciency}. In this case, we used the Perdew-Burke-Ernzerhof (PBE) generalized gradient approximation \cite{GGA} for the exchange-correlation functional without spin-orbit coupling. We employed projector augmented wave pseudopotentials \cite{PAW1,PAW2}, considering Cr $3p$ electrons as valence (Cr\_pv). The energy cutoff was set to 350~eV and we used a $15 \times 15 \times 15$ Monkhorst-Pack $k$-point mesh. We used an AMIX of $0.05$ and a SIGMA smearing parameter of $0.05$~eV. Finally, a tight-binding model is constructed using \textsc{Wannier90} \cite{Mostofi2014updated, Marzari1997maximally, Souza2001maximally}. We performed a Wannier projection of Cr $3d$ and I $5p$ orbitals without maximal localization. Disentanglement was omitted since these states form a well-isolated band set within the chosen energy window. The band structure from \gls*{DFT} calculations and the Wannier projected bands are shown in Fig.~S12. The ligand orbitals were constructed from the appropriate linear combinations of the Wannier orbitals \cite{repo}.  The hopping terms we obtained from this method are listed below (in units of eV).

\begin{widetext}
\begin{equation}\label{eq:hopping}
H_\text{hopping} = 
\begin{pNiceMatrix}[first-row,first-col]
         &d_{3z^2-r^2} &d_{xz}&d_{yz}&d_{x^2-y^2}&d_{xy}&L_{3z^2-r^2}&L_{xz}&L_{yz}&L_{x^2-y^2}&L_{xy} \\
d_{3z^2-r^2}   &-9.634 & 0.002 & 0.001  & 0      & -0.003 & -2.008 & -0.011 & -0.019 & -0.001 &  0.03  \\
d_{xz}         & 0.002 &-10.244& 0.003  & 0.003  &  0.003 & -0.001 & -1.272 & -0.001 &  0.007 & -0.002 \\
d_{yz}         & 0.001 & 0.003 &-10.244 & -0.003 &  0.003 &  0.006 & -0.002 & -1.272 & -0.002 & -0.001 \\
d_{x^2-y^2}    & 0.0   & 0.003 & -0.003 & -9.634 &  0     &  0.001 & -0.028 &  0.024 & -2.008 &  0.004 \\
d_{xy}         &-0.003 & 0.003 &  0.003 &  0     &-10.244 &  0.005 & -0.001 & -0.002 & -0.004 & -1.272 \\
L_{3z^2-r^2}   &-2.008 &-0.001 &  0.006 &  0.001 & -0.005 &  2.689 &  0.01  &  0.015 &  0.0   & -0.026 \\
L_{xz}         &-0.011 &-1.272 & -0.002 & -0.028 & -0.001 &  0.01  &  1.208 & -0.017 &  0.024 & -0.017 \\
L_{yz}         &-0.019 &-0.001 & -1.272 &  0.024 & -0.002 &  0.015 & -0.017 &  1.208 & -0.021 & -0.017 \\
L_{x^2-y^2}    &-0.001 & 0.007 & -0.002 & -2.008 & -0.004 &  0.0   &  0.024 & -0.021 &  2.689 & -0.003 \\
L_{xy}         & 0.03  &-0.002 & -0.001 &  0.004 & -1.272 & -0.026 & -0.017 & -0.017 & -0.003 &  1.208 \\
\end{pNiceMatrix}
\end{equation}
\end{widetext}

Since hopping depends on how electronic wavefunctions are spread between different atoms, it is only weakly influenced by strongly correlated physics and \gls*{DFT} generally captures the magnitude of hopping quite accurately. For this reason, we consider the off-diagonal inter-orbital hopping values fixed to the quoted \gls*{DFT}-derived values. The on-site energies correspond to the diagonal elements in the hopping matrix $H_\text{hopping}$ above. These on-site energies for the Cr $3d$ states are not easily obtainable from first principles because of the effects of strong correlations and the challenges in handling double-counting effects. We therefore consider the Cr $3d$ crystal field as a fitting parameter, which, in view of the approximately cubic symmetry of the Cr coordination, is specified by $10D_q$, which represents the splitting between the $t_{\mathrm{2g}}$ and $e_{\mathrm{g}}$ orbitals.

Since Coulomb interactions in the I $5p$ states are relatively weak and since the probability that these states are occupied by multiple holes is relatively low, the crystal field on the states is more accurately captured by \gls*{DFT}, so to reduce the number of fitting parameters, we fix the ligand orbital crystal field $10D_q^L$ to the value extracted from our \gls*{DFT} results ($1.481$~eV). In this work, we define the charge-transfer, $\Delta$, and Coulomb repulsion, $U_{dd}$, energies as the energy cost of specific transitions in the material with the Cr-ligand hopping switched off. $U_{dd}$ reflects a $d_i^3d_j^3\rightarrow d^{2}_id_j^{4}$ transition and $\Delta$ is defined as the energy for a $d_i^3\rightarrow d^{4}_i\underline{L}$ transition, where $i$ and $j$ label Cr sites and $\underline{L}$ denotes an iodine ligand hole. These energies include the effect of the Cr crystal field and represent the energies for a transition into the lowest-energy \emph{ligand} orbital (rather than the center of the I $p$ states). From a practical point of view, we perform our calculations varying the energy splitting of the Cr and ligand states and we subsequently determine $\Delta$ by diagonalizing the isolated Cr and ligand configurations and computing the appropriate differences in energy. The values of the diagonal elements in Eq.~\ref{eq:hopping} are the final onsite energies determined for our model.

In the intermediate states with the presence of a core hole, we include a core-hole potential $U_{dp}$. $U_{dd}$ is usually 4-6 eV in early transition metals and it is usually smaller when the charge transfer energy is small (which we will see is the case here), so we choose 4~eV \cite{Bocquet1992systematics, Feldkemper1998generalized}. $U_{dp}$ is only weakly dependent on solid state effects, so we choose a typical value of 6~eV.

The spin-orbit coupling terms for the Cr $3d$ orbitals ($\zeta_{\mathrm{i}}$ and $\zeta_{\mathrm{n}}$ for the initial and intermediate states, respectively) are weak and have negligible effects on the spectra. We consequently simply fixed them to their atomic values. Since we only measured \gls*{RIXS} energy map at the $L_3$-edge, we also simply fixed the core-hole spin-orbit coupling parameter $\zeta_{\mathrm{c}}$ to its atomic value. 

A small Zeeman interaction term $g \mu_B \mathbf{B} \cdot \mathbf{S}$ was applied to the total spin angular momentum of the system, serving as the effective exchange field in the magnetically ordered state. We fixed $\mu_B B=0.005$~eV to match the energy scale of the exchange interactions in \ce{CrI3} \cite{Chen2018Topological}.

In summary, we have four free parameters in our model, i.e., $k_{dd}$, $k_{dp}$, $10D_q$, and $\Delta$. Most of these values are constrained by physical considerations. $k_{dd}$ and $k_{dp}$ quantify the screening. These are known to vary within a range of 0.5--0.9 from prior studies and these values will typically also have reasonably similar values since both are affected by similar screening processes \cite{Groot2008core}. $10D_q$ in $3d$ octahedrally coordinated transition metal material ranges from about 0.5--3~eV and since iodine is relatively large size in ionic radius and relativity weakly electronegative, we expect a value in the lower half of this range. $k_{dd}$, $k_{dp}$, $10D_q$, and $\Delta$ have distinct effects on the \gls*{RIXS} energy map. $k_{dd}$ directly scales the Hund's coupling and hence controls the energies of the dark excitons D1 and D2. The energies of the bright excitons B1 and B2 are primarily determined by both $10D_q$ and $\Delta$. $k_{dp}$ mainly affects the resonance profiles of these peaks. Thanks to the richly detailed spectra, the finite physically reasonable range of parameters, and the distinct effects of different parameters, we successfully obtained a well-constrained model with final parameters as $k_{dd}=0.65$, $k_{dp}=0.5$, $10D_q = 0.61$~eV, $\Delta = -1.3$~eV and verified that no other solutions exist. The level of agreement compares favorably with what can be expected for a model of this type accurately capturing the energy of the excitations while roughly capturing trends in peak intensities \cite{Kang2020Coherent, Son2022multiferroic, He2024Magnetically, Occhialini2024Nature}. The full list of parameters used in the model is shown in Tab.~\ref{table:ED_AIM}. The estimated error bars are $\sim 1$~eV for all the Coulomb interactions and the charge-transfer energy $\Delta$ and $\sim 0.2$~eV for $10D_q$.

\begin{table*}
\caption{\textbf{Full list of parameters used in the \gls*{AIM} calculations}. (Except for the hopping integrals, which are provided in Eq.~\ref{eq:hopping}.) $F_{dd,\mathrm{i}}$ and $F_{dd,\mathrm{n}}$ are for the initial and intermediate states, respectively. Units are eV.}
\begin{ruledtabular}
\begin{tabular}{cccccccccccccccccc}
$10D_q$ & $10D_q^L$ & $\Delta$ & $U_{dd}$ & $U_{dp}$ & $F^2_{dd,\mathrm{i}}$ & $F^4_{dd,\mathrm{i}}$ & $F^2_{dd,\mathrm{n}}$ & $F^4_{dd,\mathrm{n}}$ & $F^2_{dp}$ & $G^1_{dp}$ & $G^3_{dp}$ & $\zeta_{\mathrm{i}}$ & $\zeta_{\mathrm{n}}$ & $\zeta_{\mathrm{c}}$ \\
\hline
0.61 & 1.481 & -1.3 & 4.0 & 6.0 & 7.005 & 4.391 & 7.537 & 4.726 & 3.263 & 2.394 & 1.361 & 0.035 & 0.047 & 5.667\\
\end{tabular}
\end{ruledtabular}
\label{table:ED_AIM}
\end{table*}

\bibliography{refs}
\end{document}


\title{Supplemental Material: Dispersive Dark Excitons in van der Waals Ferromagnet CrI$_3$}

\renewcommand{\thesection}{S\arabic{section}}  
\renewcommand{\thetable}{S\arabic{table}}  
\renewcommand{\thefigure}{S\arabic{figure}}

\date{\today}

\maketitle

This document provides a \gls*{RIXS} energy map measured with linear vertical ($\sigma$) polarization, \gls*{XAS} data, additional temperature dependence of \gls*{RIXS} spectra measured at $E_{e_{\mathrm{g}}}$ resonance, detailed description of the fitting procedures for the \gls*{RIXS} data with linecuts showing the fitting, plots showing integrated peak intensity temperature dependence, further details of our \gls*{ED} calculations, and a discussion of terminology for the excitations studies here.

\section{Additional \gls*{RIXS} energy map with $\sigma$ polarization and \gls*{XAS} data}\label{sec:sigmapol}
\begin{figure}
\includegraphics{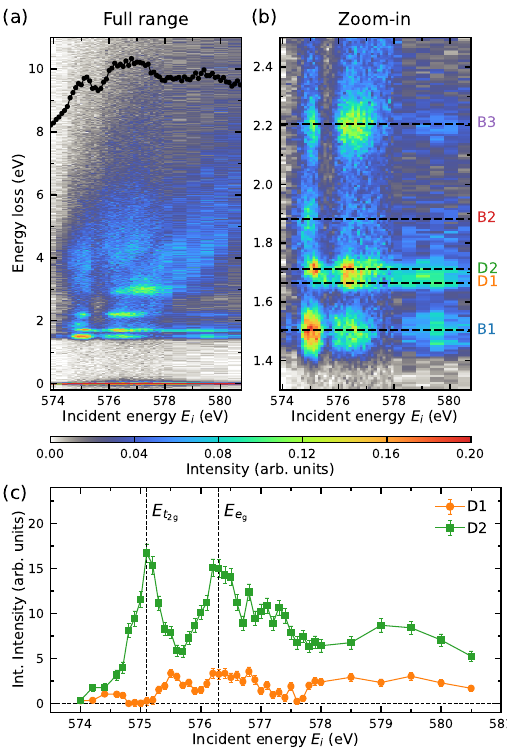}
\caption{Additional resonance behavior of dark excitons measured with linear vertical ($\sigma$) polarization. (a) Incident photon energy $E_i$ dependence of the \gls*{RIXS} intensity map at the Cr $L_3$-edge with $\sigma$-polarized incident x-rays. The measurement was taken at $T=30$~K with x-rays incident on the sample at $\theta = 14.5^\circ$ and scattered to $2\Theta = 150^\circ$ in the ($H0L$) scattering plane. The overlaid black curve on the top is \gls*{XAS} spectrum taken at the same conditions (including x-ray polarization, experimental geometry, and temperature). (b) Zoom-in of the \gls*{RIXS} energy map in the vicinity of the dark excitons. Horizontal dashed lines are fitted exciton energies. (c) The fitted integrated intensities of the two dark excitons as a function of incident photon energy $E_i$. Error bars represent one standard deviation. Similar to the $\pi$ polarization data in the main text, both dark excitons reach their local maxima in intensity at $E_{e_{\mathrm{g}}}$, while D1 has nearly zero spectral weight at $E_{t_{\mathrm{2g}}}$. 
}
\label{fig:SI_energy_map_LV}
\end{figure}

\begin{figure}
\includegraphics{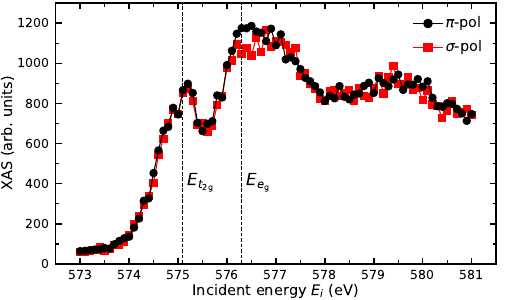}
\caption{XAS spectra with both linear horizontal ($\pi$) and linear vertical ($\sigma$) polarizations. The spectra were taken at $T=30$~K. The sample geometry was the same as the measurement for the RIXS energy map (i.e., $\theta = 14.5^\circ$, $2\Theta = 150^\circ$, ($H0L$) scattering plane) but with a much larger exit slit size of $300$~{\textmu}m. The vertical dashed lines label the two resonant energies. This is the same data shown in Figs.~1 and \ref{fig:SI_energy_map_LV} and provided here to show the comparison between the two polarizations.
}
\label{fig:SI_XAS}
\end{figure}

Besides the incident energy dependent \gls*{RIXS} intensity map measured with linear horizontal ($\pi$) polarization shown in the main text, we also collected data with linear vertical ($\sigma$) polarization as displayed in Fig.~\ref{fig:SI_energy_map_LV}. All experimental parameters were kept the same for these two sets of measurements, with the only difference being the incident x-ray polarization. The two \gls*{RIXS} energy maps are very similar with nearly identical resonance behaviors, although the $\sigma$-channel data has slightly lower intensity. We therefore chose $\pi$-polarization for the following angular and temperature-dependent measurements to maximize the \gls*{RIXS} intensity. The \gls*{XAS} also exhibits a weak dichroism for the two polarizations, as shown in Fig.~\ref{fig:SI_XAS}.

\section{Additional temperature dependence of \gls*{RIXS} spectra at $E_{e_{\mathrm{g}}}$ resonance}

\begin{figure}
\includegraphics{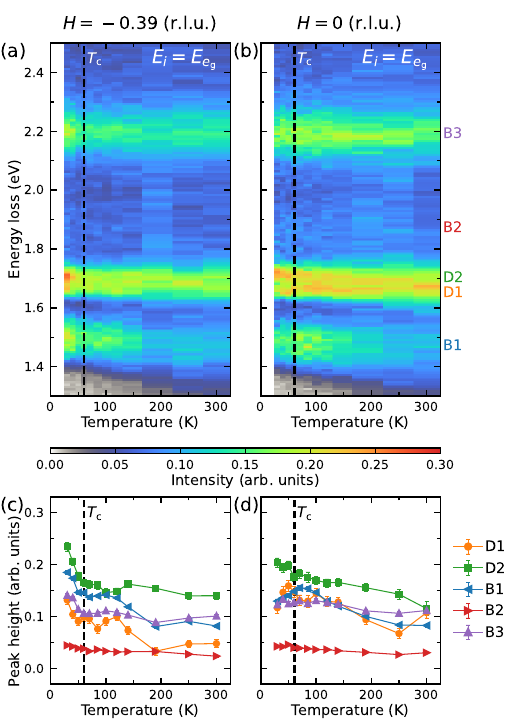}
\caption{Additional temperature dependence of the dark excitons at $E_{e_{\mathrm{g}}}$ resonance. (a),(b) \gls*{RIXS} intensity map as a function of temperature measured at two different in-plane momenta ($H=-0.39$ and $H=0$ r.l.u., respectively) with the same incident energy $E_{e_{\mathrm{g}}}$. (c),(d) Corresponding peak height extracted from the fits as a function of temperature. Error bars represent one standard deviation. The dashed lines indicate the \gls*{FM} transition temperature $T_{\mathrm c}$.
}
\label{fig:SI_Tdep}
\end{figure}

Besides the temperature dependent \gls*{RIXS} intensity map measured at $E_{t_{\mathrm{2g}}}$ resonance shown in the main text, we also have data collected at $E_{e_{\mathrm{g}}}$ resonance as displayed in Fig.~\ref{fig:SI_Tdep}. As described in detail in the next section, we co-fit these two data sets with shared exciton widths and energies. Just like the data at $E_{t_{\mathrm{2g}}}$, the excitons also exhibit abrupt changes in peak height near the \gls*{FM} transition temperature $T_{\mathrm c}$ at this resonance, although the changes are less dramatic.

\section{Fitting of the \gls*{RIXS} spectra}
\subsection{Functional form}
We fit \gls*{RIXS} spectra with five peaks plus a linear background to parameterize the excitons as a function of incident energy, momentum transfer, or temperatures. We used Voigt functions to fit the dark excitons D1--D2 and bright excitons B2--B3. The width of the Gaussian component was fixed to the energy resolution of $30.5$~meV, which was determined by a reference measurement on a multilayer heterostructure sample with a strong elastic scattering signal. The bright exciton B1 is better described by a Gaussian lineshape. We found that these simple lineshapes provide a satisfactory fit to the spectra. 

To obtain high-precision values for the exciton energies, we carefully calibrated the energy zero of the spectra. In the low-energy region of the spectra, we identified one prominent elastic peak along with a weak magnon feature in the tail of peak. We used a Gaussian function with its width fixed to the energy resolution to fit the elastic peak. For the magnon, we chose a \gls*{DHO} function below and convoluted this function with a Gaussian resolution function to model it. The form of the \gls*{DHO} function we used is:
\begin{equation}
    S(\Q{},\omega)=\frac{\omega \chi_Q}{1-\exp(-\omega/k_B T)}\cdot \frac{2z_Q f_Q}{(\omega^2-f_Q^2)^2+(\omega z_Q)^2}
\end{equation}
where $f_Q$ is the undamped energy, $\chi_Q$ is the oscillator strength, $z_Q$ is the damping factor, $k_B$ is the Boltzmann constant, and $T$ is temperature. Because the elastic peak dominates the spectral weight in this region, the inclusion of the magnon peak in the fitting function has minimal effect on the fitted elastic peak position. The error
bars for the fitted exciton peak positions shown in Fig.~2 include not only their own fitting errors but also the fitting errors of the energy zero.

\subsection{Incident energy dependence}
\begin{figure}
\includegraphics{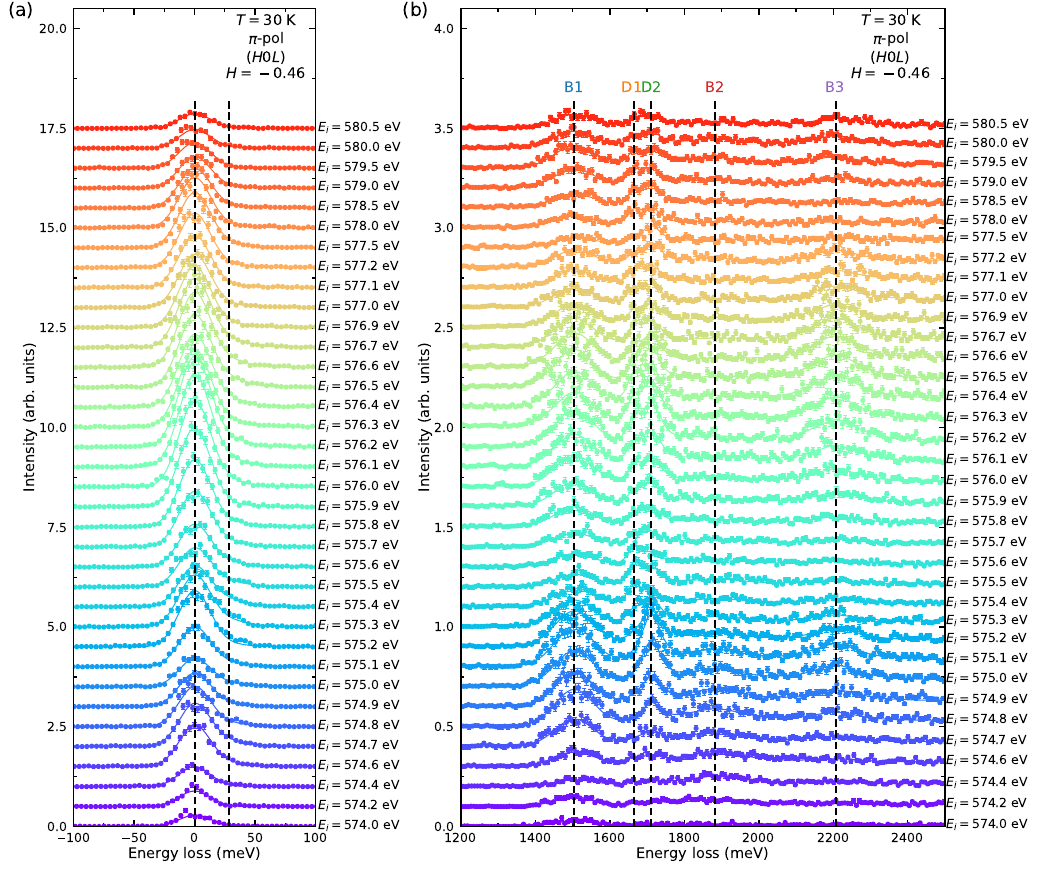}
\caption{\gls*{RIXS} spectra measured at various incident energies with linear-horizontally ($\pi$-) polarized x-rays. (a) \gls*{RIXS} spectra near the elastic line. (b) \gls*{RIXS} spectra in the energy window covering the excitons. The measurements were taken at various incident photon energy through the Cr $L_3$-edge measured in the ($H0L$) plane at $T = 30$~K. These data are the same as the intensity maps shown in Fig.~1(a)--(b) and are provided to show the linecuts directly. Data are shifted vertically for clarity. The solid lines are the fits to the data. Error bars represent one standard deviation. Vertical black dashed lines label the exciton energies.
}
\label{fig:SI_energy_map_LH_fits}
\end{figure}

\begin{figure}
\includegraphics{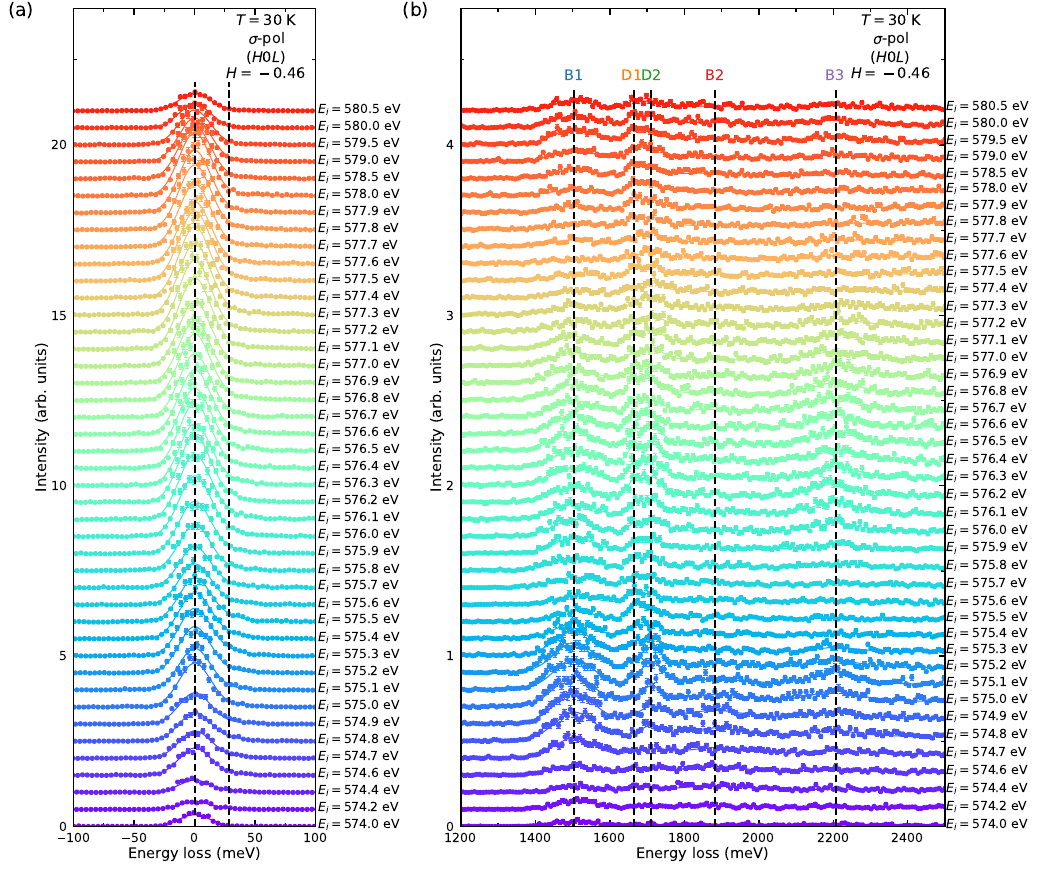}
\caption{\gls*{RIXS} spectra measured at various incident energies with linear-vertically ($\sigma$-) polarized x-rays. (a) \gls*{RIXS} spectra near the elastic line. (b) \gls*{RIXS} spectra in the energy window covering the excitons. The measurements were taken at various incident photon energy through the Cr $L_3$-edge measured in the ($H0L$) plane at $T = 30$~K. These data are the same as the intensity maps shown in Fig.~\ref{fig:SI_energy_map_LV}(a)--(b) and are provided to show the linecuts directly. Data are shifted vertically for clarity. The solid lines are the fits to the data. Error bars represent one standard deviation. Vertical black dashed lines label the exciton energies.
}
\label{fig:SI_energy_map_LV_fits}
\end{figure}

As expected for a Raman-like \gls*{RIXS} process, the exciton energies and widths were found to be independent of incident energy, providing additional constraints in fitting the incident energy dependent \gls*{RIXS} spectra. Specifically, we first co-fitted all the spectra taken with $\pi$-polarization, applying the constraints that each exciton maintains the same position and width across all spectra (see Fig.~\ref{fig:SI_energy_map_LH_fits}). Then, we co-fitted all the spectra taken with $\sigma$-polarization. As shown in Fig.~\ref{fig:SI_energy_map_LV_fits}, we found that even when we forced the exciton positions and widths to match the fitted values obtained from the $\pi$-polarization data, the fits still look reasonably good, demonstrating the robustness of the fitting result. 

\subsection{In-plane momentum transfer dependence}
\begin{figure}
\includegraphics{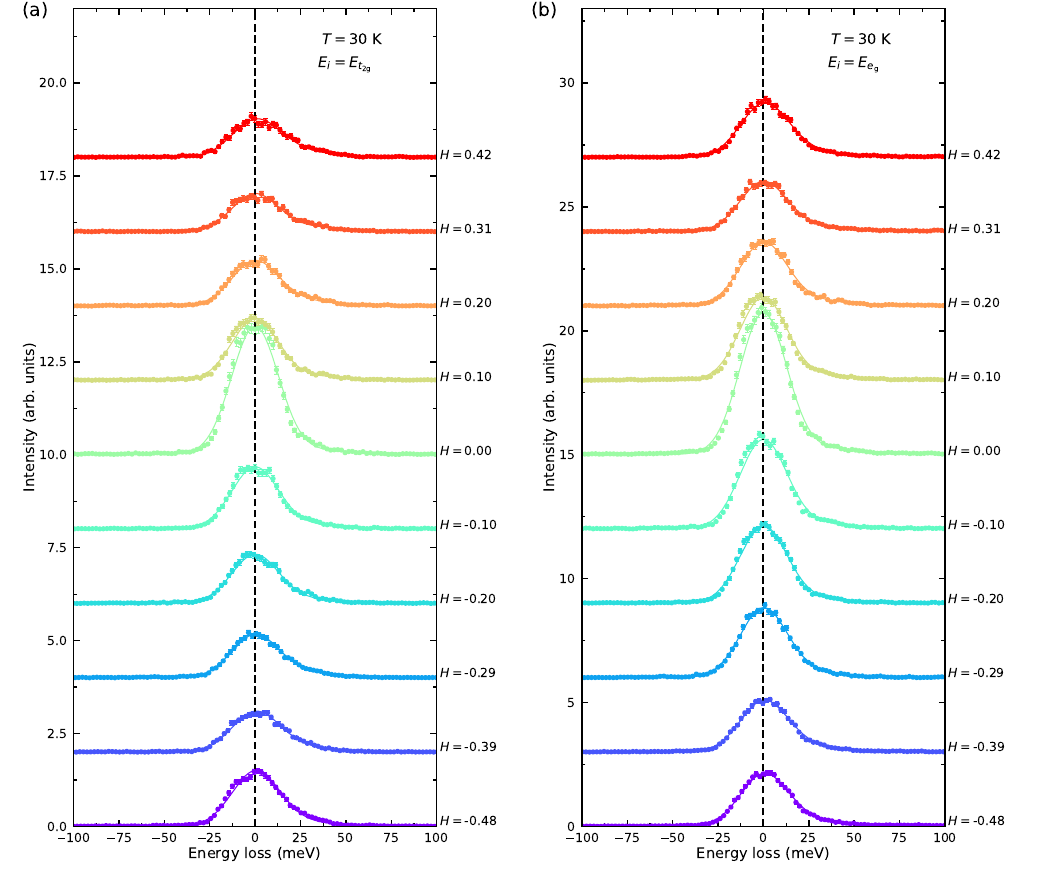}
\caption{\gls*{RIXS} spectra measured at various in-plane momentum transfer $H$ with an energy window chosen to isolate the elastic line. (a) \gls*{RIXS} spectra at $E_{t_{\mathrm{2g}}}$ resonance. (b) \gls*{RIXS} spectra at $E_{e_{\mathrm{g}}}$ resonance. The measurements were taken in the ($H0L$) plane at $T = 30$~K with linear horizontal ($\pi$) polarization of the incident x-rays. Data are shifted vertically for clarity. The solid lines are the fits to the data. Error bars represent one standard deviation. Vertical black dashed lines label the energy zero.
}
\label{fig:SI_dispersion_elastic_fits}
\end{figure}

\begin{figure}
\includegraphics{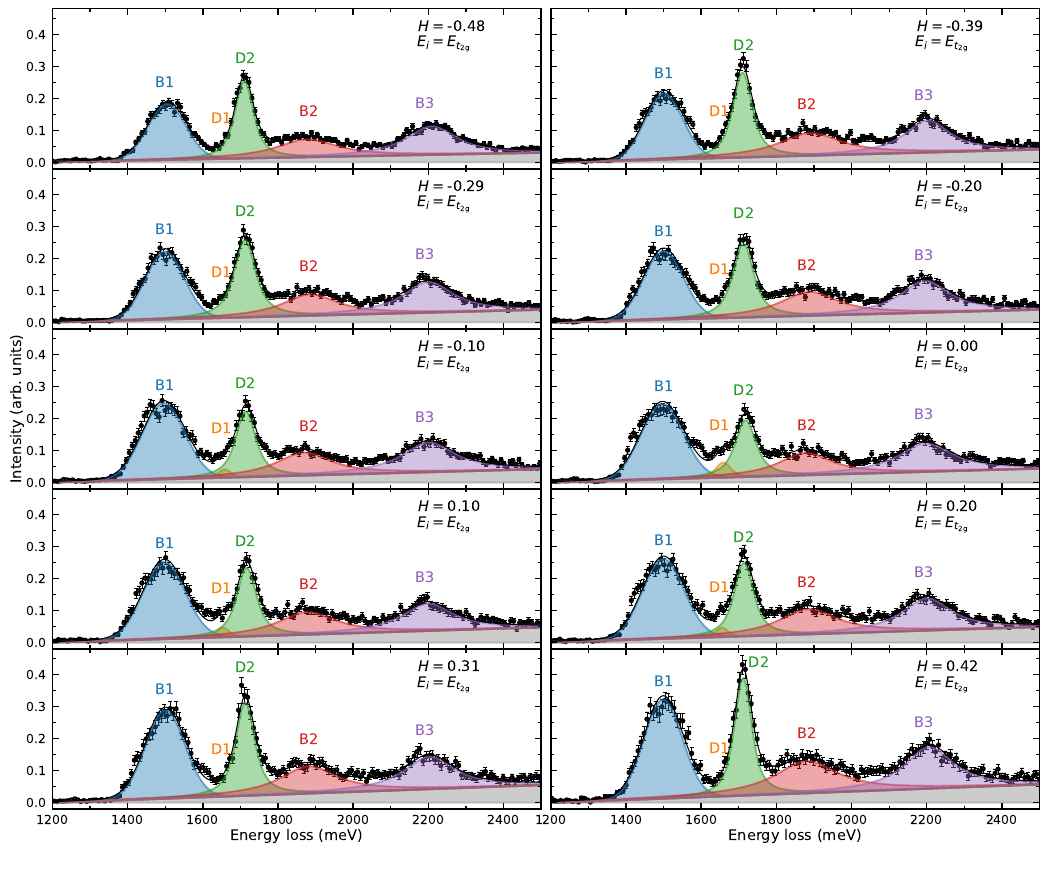}
\caption{\gls*{RIXS} spectra measured at various in-plane momentum transfer $H$ at $E_{t_{\mathrm{2g}}}$ resonance with an energy window chosen to isolate the excitons. The measurements were taken in the ($H0L$) plane at $T = 30$~K with linear horizontal ($\pi$) polarization of the incident x-rays. These data are the same as the intensity maps shown in Fig.~2(a) and are provided to show the linecuts directly. Solid black lines are the fits to the data, with shaded areas showing the contributions from different components. Error bars represent one standard deviation. 
}
\label{fig:SI_dispersion_excitons_fits_Et2g}
\end{figure}

\begin{figure}
\includegraphics{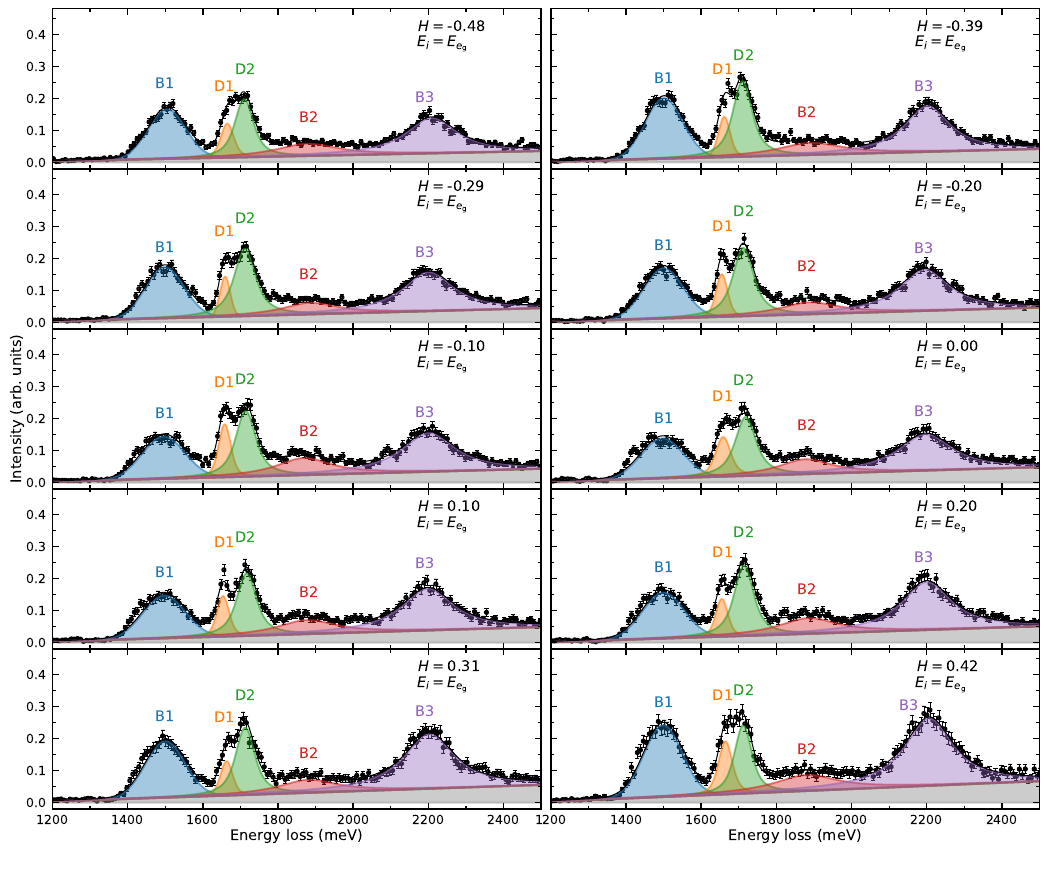}
\caption{\gls*{RIXS} spectra measured at various in-plane momentum transfer $H$ at $E_{e_{\mathrm{g}}}$ resonance with an energy window chosen to isolate the excitons. The measurements were taken in the ($H0L$) plane at $T = 30$~K with linear horizontal ($\pi$) polarization of the incident x-rays. These data are the same as the intensity maps shown in Fig.~2(b) and are provided to show the linecuts directly. Solid black lines are the fits to the data, with shaded areas showing the contributions from different components. Error bars represent one standard deviation. 
}
\label{fig:SI_dispersion_excitons_fits_Eeg}
\end{figure}

\begin{figure}
\includegraphics{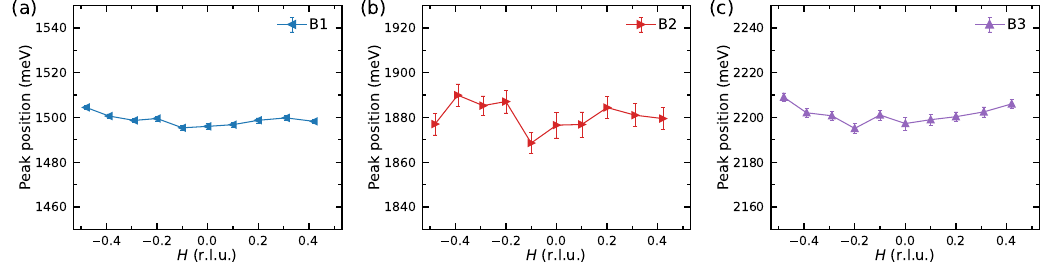}
\caption{The extracted momentum dependent energies of the three bright excitons. Error bars represent one standard deviation. No clear dispersion is detected for these bright excitons.
}
\label{fig:SI_dispersion_bright_excitons}
\end{figure}

For the fitting of the \gls*{RIXS} spectra at various in-plane momentum transfers, since we have measurements at two incident energies, we applied the above-mentioned co-fitting method and constrained the exciton energy and widths to be the same across both spectra. Such a co-fitting strategy with shared exciton positions and widths is particularly useful in obtaining consistent fits for the dark exciton D1, which has a minuscule spectral weight at $E_{t_{\mathrm{2g}}}$ resonance. We present the best fits to the linecuts of the RIXS spectra in Fig.~\ref{fig:SI_dispersion_elastic_fits}--\ref{fig:SI_dispersion_excitons_fits_Eeg}. In Fig.~\ref{fig:SI_dispersion_bright_excitons}, we show the fitted peak positions of the three bright excitons as a function of the in-plane momentum transfers. No clear dispersion is detected for these excitons beyond their error bars.

\subsection{Temperature dependence}
\label{sec:fitting_Tdep}
\begin{figure}
\includegraphics{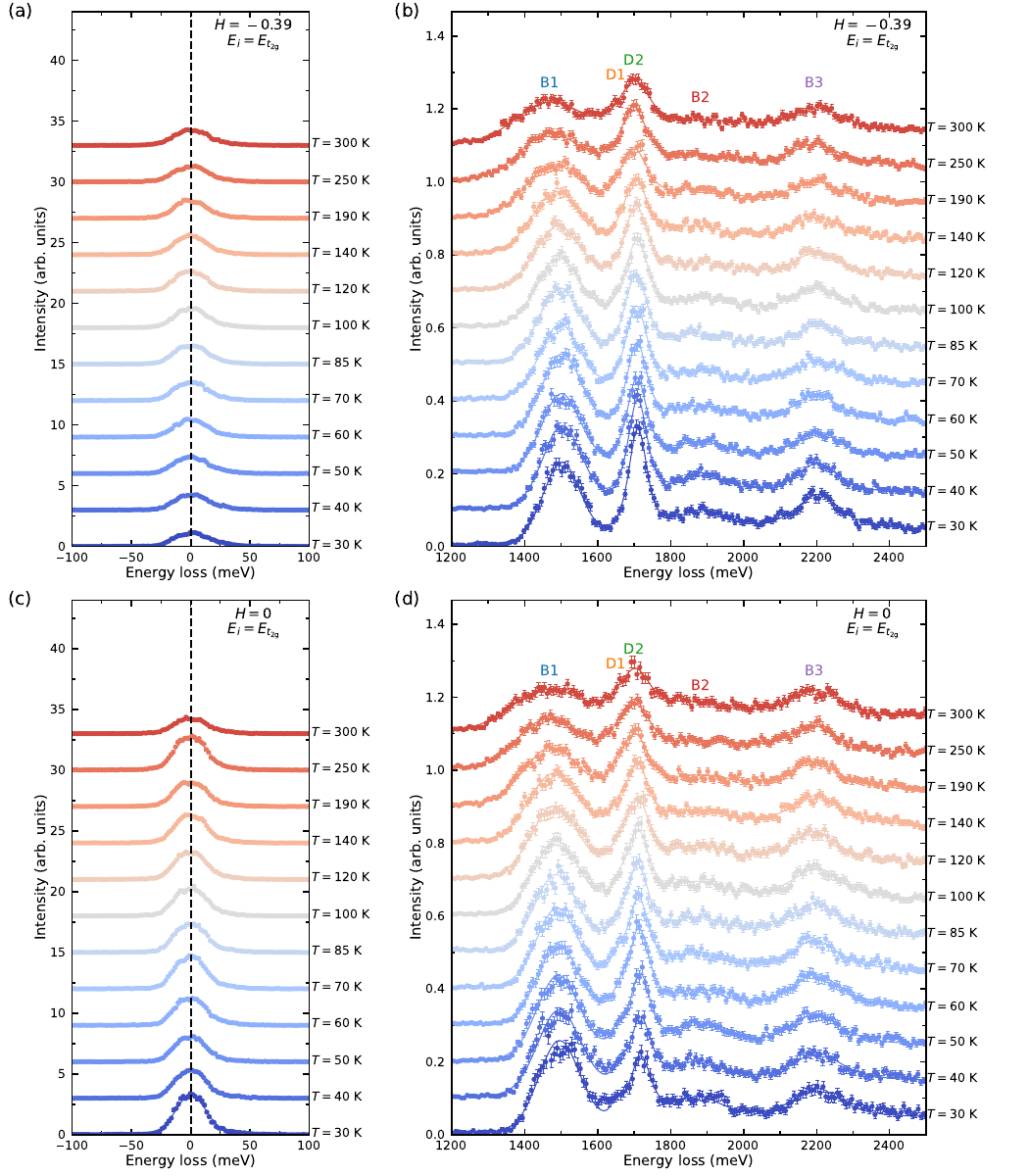}
\caption{\gls*{RIXS} spectra measured at various temperatures at $E_{t_{\mathrm{2g}}}$ resonance. (a),(b) \gls*{RIXS} spectra taken at $H=-0.39$ r.l.u.\ with the energy window chosen to isolate the elastic line and excitons, respectively. (c),(d) \gls*{RIXS} spectra taken at $H=0$ r.l.u.\ with the energy window chosen to isolate the elastic line and excitons, respectively. All the measurements were done with linear horizontal ($\pi$) polarization of the incident x-rays. These data are the same as the intensity maps shown in Fig.~3(a)--(b) and are provided to show the linecuts directly. Data are shifted vertically for clarity. The solid lines are the fits to the data. Error bars represent one standard deviation. Vertical dashed lines label the energy zero.
}
\label{fig:SI_Tdep_fits_Et2g}
\end{figure}

\begin{figure}
\includegraphics{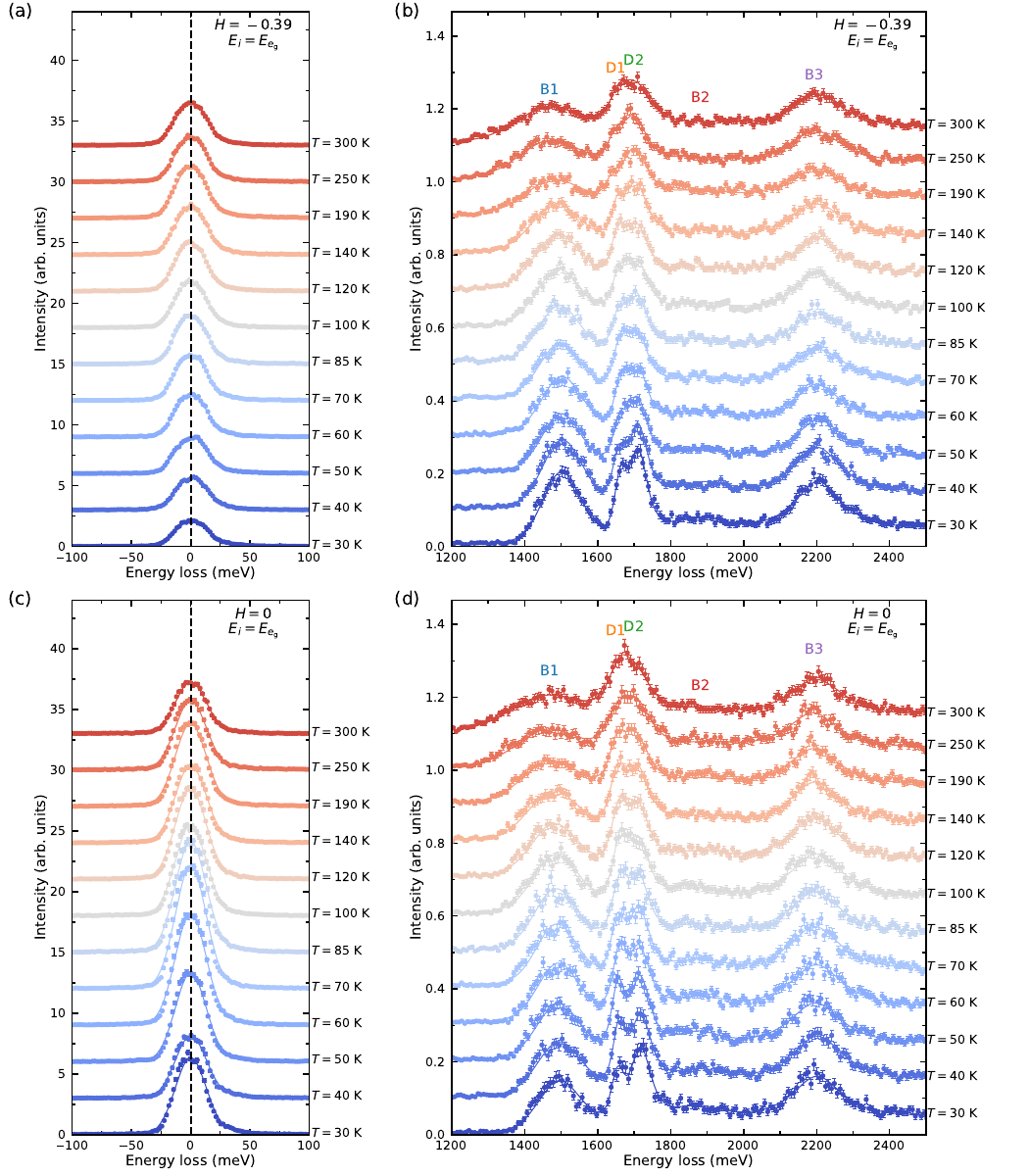}
\caption{\gls*{RIXS} spectra measured at various temperatures at $E_{e_{\mathrm{g}}}$ resonance. (a),(b) \gls*{RIXS} spectra taken at $H=-0.39$ r.l.u.\ with the energy window chosen to isolate the elastic line and excitons, respectively. (c),(d) \gls*{RIXS} spectra taken at $H=0$ r.l.u.\ with the energy window chosen to isolate the elastic line and excitons, respectively. All the measurements were done with linear horizontal ($\pi$) polarization of the incident x-rays. These data are the same as the intensity maps shown in Fig.~\ref{fig:SI_Tdep}(a)--(b) and are provided to show the linecuts directly. Data are shifted vertically for clarity. The solid lines are the fits to the data. Error bars represent one standard deviation. Vertical dashed lines label the energy zero.
}
\label{fig:SI_Tdep_fits_Eeg}
\end{figure}

\begin{figure}
\includegraphics{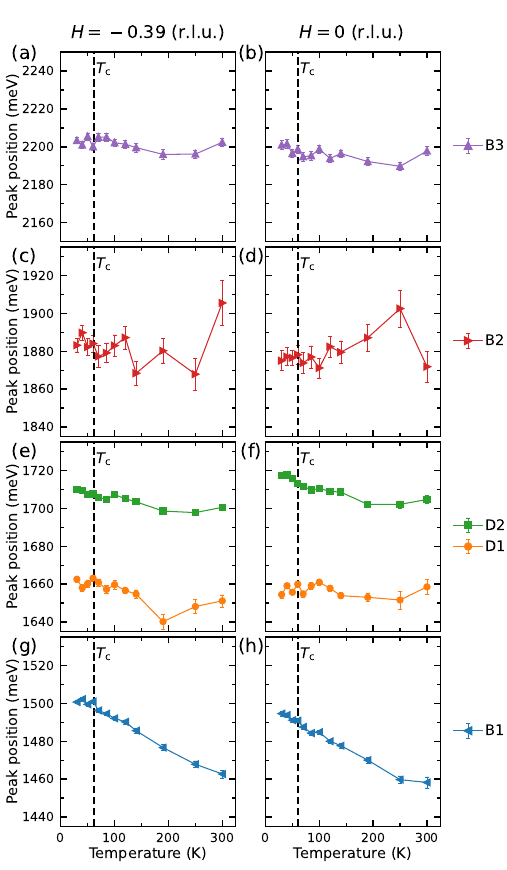}
\caption{The extracted temperature dependent energies of the excitons at two different in-plane momenta ($H=-0.39$ and $H=0$ r.l.u.). Error bars represent one standard deviation. The dashed lines indicate the \gls*{FM} transition temperature $T_{\mathrm c}$.
}
\label{fig:SI_Tdep_center}
\end{figure}

We used the same co-fitting approach to fit the temperature-dependent \gls*{RIXS} spectra. However, due to the broadening of the exciton peaks, not every spectrum yielded a converged fitting result. Therefore, we added additional constraints, including the integrated intensity of B2 at both resonance energies and the integrated intensity of D1 at the $E_{t_{\mathrm{2g}}}$ resonance. We found that these quantities have minimal temperature dependence, so we fixed them to their respective average values obtained from the initial fits. The final best fits to the temperature-dependent RIXS spectra are shown in Fig.~\ref{fig:SI_Tdep_fits_Et2g} and \ref{fig:SI_Tdep_fits_Eeg}. Figure \ref{fig:SI_Tdep_center} displays the fitted peak positions of all five excitons as a function of temperature measured. B1 (and to a lesser extent D2) shows a clear red-shift upon warming. The red shift is less apparent for other excitons, partly due to their weak intensities and/or broad peak width.

\clearpage

\section{Integrated peak intensity temperature dependence}
\begin{figure}
\includegraphics{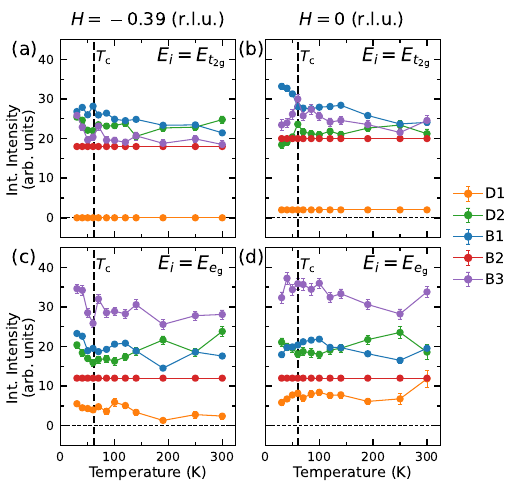}
\caption{Temperature dependence of the dark excitons at (a),(c) $H = -0.39$~r.l.u. and (b),(d) $H = 0$ at two different in-plane momenta ($H=-0.39$ and $H=0$ r.l.u.). This re-plots the same data as in Fig.~3 of the main text and Fig.~\ref{fig:SI_Tdep}, but using integrated intensity rather than peak height. Error bars represent one standard deviation.}
\label{fig:SI_Tdep_amplitude}
\end{figure}

For completeness, here in Fig.~\ref{fig:SI_Tdep_amplitude} we re-plot the same data provided in Fig.~3 of the main text and Fig.~\ref{fig:SI_Tdep} in terms of integrated peak intensity rather than peak height. The anomalies across $T_{\mathrm c}$ are apparent in integrated peak intensity as well. This is because of the slow change in peak width below $T_{\mathrm c}$. Note that the integrated intensity of B2 at both resonance energies and and D1 at the $E_{t_{\mathrm{2g}}}$ resonance are fixed in the whole temperature range as explained in Sec.~\ref{sec:fitting_Tdep}.

\clearpage

\section{Further details of ED RIXS calculations}
\begin{figure}
\includegraphics[width=0.45\textwidth]{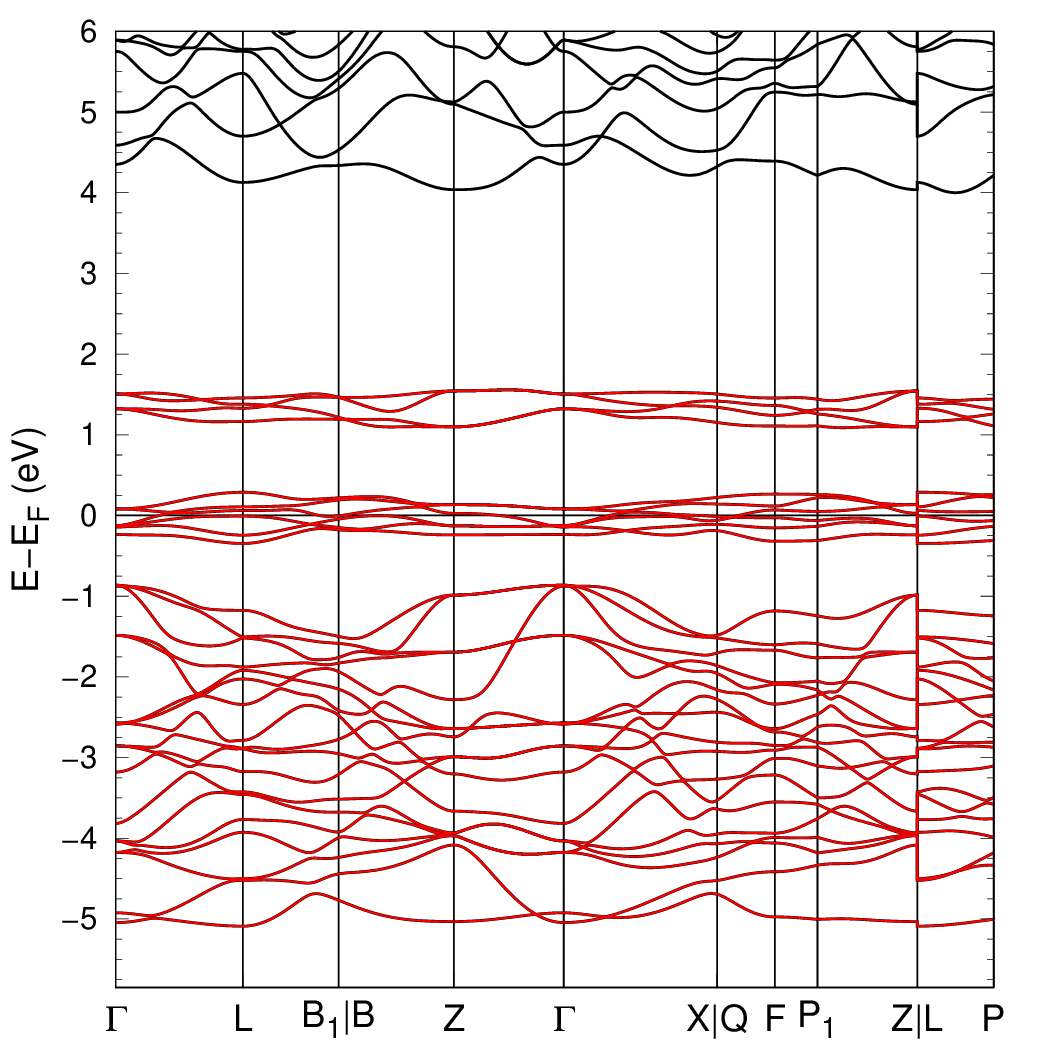}
\caption{Band structure of \ce{CrI3} without spin-orbit coupling. Wannier projected bands are in red, overlain on the \gls*{DFT} calculated bands in black showing their agreement.}
\label{fig:SI_bandstructure}
\end{figure}

To verify the suitability of the hopping matrix in the main text, we overlay the Wannier projected bands with the original \gls*{DFT} in Fig.~\ref{fig:SI_bandstructure}. We also provide further details of our atomic model for \ce{CrI3} and the spin and angular dependence of the spectra.

\subsection{Single site atomic model}\label{ref:atomic_model}
\begin{figure}
\includegraphics{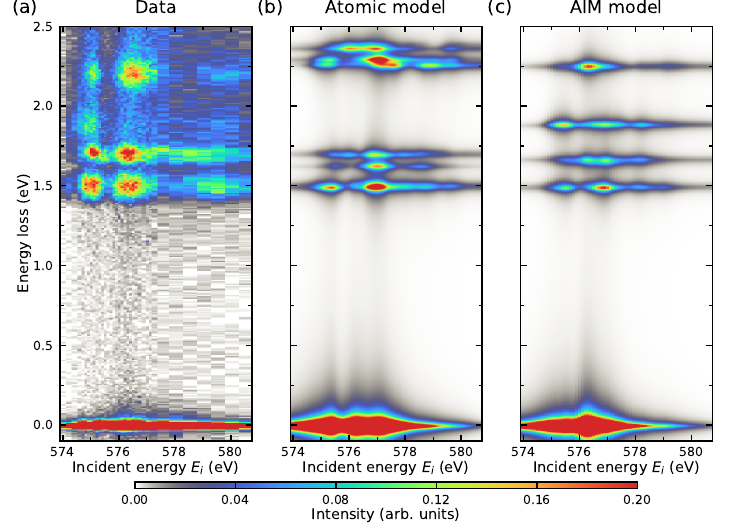}
\caption{Comparison of \gls*{RIXS} energy maps calculated using different models. (a) The measured \gls*{RIXS} energy map. (b) The calculated \gls*{RIXS} energy map using an atomic model. (c) The calculated \gls*{RIXS} energy map using an \gls*{AIM}. Panels (a),(c) use the same data shown in Fig.~4(a)--(b) and are replotted here for comparison. 
}
\label{fig:SI_ED}
\end{figure}

We also tested a simpler model that only considers a single \ce{Cr^3+} ion with 10 effective $3d$ orbitals. In this case, only three independent parameters are used, including $k_{dd}$, $k_{dp}$ and $10D_q$. Again, $k_{dd}$ dictates the energies of the dark excitons D1 and D2, and $k_{dp}$ affects the resonance profiles of the excitations. Therefore, their refined values ($k_{dd}=0.61$, $k_{dp}=0.47$) are very close to that in the \gls*{AIM}. However, the energy of exciton B1 is exclusively governed by $10D_q$, which gives $10D_q = 1.5$~eV. However, this choice of $10D_q$ limits the energy of exciton B2 to $\sim 2.3$~eV, much higher than the experimental value of $1.88$~eV. Therefore, without explicitly including the ligand orbitals, we could not simultaneously obtain the correct energies for both B1 and B2. A comparison between the two models is displayed in Fig.~\ref{fig:SI_ED}, with the full parameters presented in Tab.~\ref{table:ED_atomic}. The failure of the atomic model further underlines the fact that \ce{CrI3} is a self-doped charge transfer material. 

\begin{table*}
\caption{\textbf{Full list of parameters used in the single site atomic model calculations}. $F_{dd,\mathrm{i}}$ and $F_{dd,\mathrm{n}}$ are for the initial and intermediate states, respectively. Units are eV.}
\begin{ruledtabular}
\begin{tabular}{cccccccccccccccccc}
$10D_q$ & $F^2_{dd,\mathrm{i}}$ & $F^4_{dd,\mathrm{i}}$ & $F^2_{dd,\mathrm{n}}$ & $F^4_{dd,\mathrm{n}}$ & $F^2_{dp}$ & $G^1_{dp}$ & $G^3_{dp}$ & $\zeta_{\mathrm{i}}$ & $\zeta_{\mathrm{n}}$ & $\zeta_{\mathrm{c}}$ \\
\hline
1.50 & 6.574 & 4.121 & 7.074 & 4.435 & 3.067 & 2.250 & 1.279 & 0.035 & 0.047 & 5.667\\
\end{tabular}
\end{ruledtabular}
\label{table:ED_atomic}
\end{table*}

\subsection{\gls*{XAS} state analysis}

\begin{figure}
\includegraphics{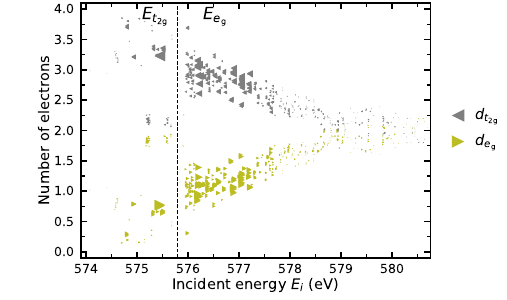}
\caption{Electron occupations of $t_{\mathrm{2g}}$ and $e_{\mathrm{g}}$ orbitals in the intermediate state of the atomic model. The symbol size (area) is proportional to the calculated intensity of each state.
}
\label{fig:SI_ED_XAS_analysis}
\end{figure}

We use $E_{t_{\mathrm{2g}}}$ and $E_{e_{\mathrm{g}}}$ to label the resonance energies at $E_i =575.1$~eV and $576.3$~eV. To support this assignment, we calculated the relevant quantum numbers for the intermediate state in the atomic model approximation, which captures the resonance behavior reasonably well as shown in Fig.~\ref{fig:SI_ED}. The result is presented in Fig.~\ref{fig:SI_ED_XAS_analysis}. As expected, there is strong mixing between $t_{\mathrm{2g}}$ and $e_{\mathrm{g}}$ orbitals, however, a substantial change in $t_{\mathrm{2g}}$ vs. $e_{\mathrm{g}}$ is seen, confirming that our labels are reasonable. 

\subsection{Angular dependence}
\begin{figure}
\includegraphics{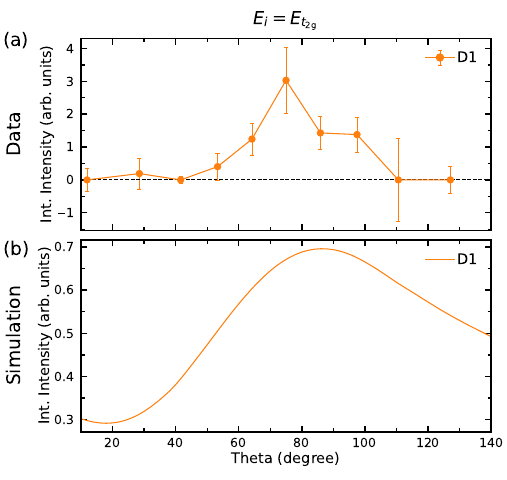}
\caption{Comparison between the measured and calculated incident-angle theta dependent dark exciton D1 intensity at $E_{t_{\mathrm{2g}}}$ resonance. The calculation is based on the \gls*{AIM}. The calculation nicely reproduces the trend of the intensity.
}
\label{fig:SI_ED_thdep}
\end{figure}

In the measured in-plane momentum dependence of the \gls*{RIXS} spectra, we observe that the intensity of the D1 exciton reaches its maximum near the Brillouin zone center at $E_{t_{\mathrm{2g}}}$ resonance. This trend is qualitatively captured by the angular dependence of the calculated \gls*{RIXS} spectra. In Fig.~\ref{fig:SI_ED_thdep}, we show the calculated intensity of this exciton D1 as a function of the incident angle $\theta$ at $E_{t_{\mathrm{2g}}}$ resonance using the \gls*{AIM}. By varying the incident angle, we essentially change the polarization of the incident and scattered photons, which ultimately affects the \gls*{RIXS} cross section. 

\subsection{Spin direction dependence}
\begin{figure}[H]
\includegraphics{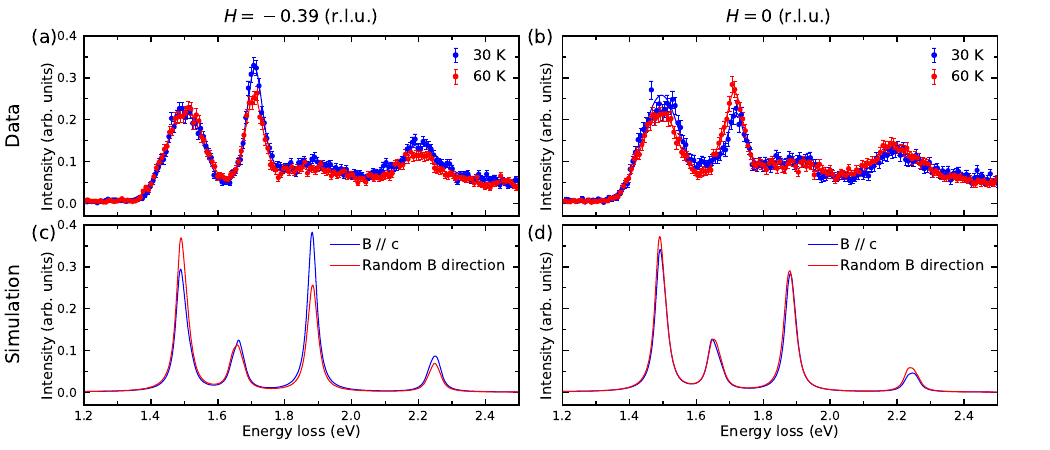}
\caption{Comparison between the measured and calculated \gls*{RIXS} spectra. (a),(b) Representative measured \gls*{RIXS} spectra at two different momentum transfers at two temperatures. The solid lines are the fits to the data. Blue data is within the magnetically ordered phase well below the \gls*{FM} transition temperature $T_{\mathrm c}$, whereas red data is near $T_{\mathrm c}$ with much reduced ordered moment. (c),(d) Corresponding calculated \gls*{RIXS} spectra based on the \gls*{AIM} to simulate the spin direction effect. The blue lines assume the spin directions are aligned along the $c$-axis, while the red lines assume random spin directions. The calculations predict negligible changes in the intensities of the dark excitons, in contrast to the experimental observations.
}
\label{fig:SI_ED_spin_direction_dep}
\end{figure}

We have shown an abrupt change in the dark exciton intensity across the \gls*{FM} transition temperature $T_{\mathrm c}$. Here, we examine one possible origin by considering the spin-direction dependent \gls*{RIXS} spectra. The rationale is that the spin direction, which is clearly different across $T_{\mathrm c}$, can potentially affect the \gls*{RIXS} cross section. Below $T_{\mathrm c}$, \ce{CrI3} is in the \gls*{FM} state with spins pointing along $c$-axis, whereas above $T_{\mathrm c}$, it enters the paramagnetic state with randomly oriented spins. This change in spin direction can be simulated by controlling the direction of the external magnetic field $\mathbf{B}$ in the Zeeman term. The representative calculated spectra based on the \gls*{AIM} are shown in Fig.~\ref{fig:SI_ED_spin_direction_dep}. The calculated dark exciton intensity barely changes in comparison to the bright excitons, contrary to the trends observed from the experimental spectra. Thus, the peculiar temperature dependence cannot be simply explained by the spin-direction dependence and likely requires more complicated explanations involving exciton-magnon interactions. 

\section{An aside on terminology}
Here, we further discuss whether the excitations in \ce{CrI3} should be classified as $dd$-excitations or, more generally, as excitons. This requires a precise definition of $dd$-excitations. The strictest definition involves transitions that, to a good approximation, occur only within local $d$-orbitals. A broader but still rigorous definition describes $dd$-excitations as transitions within an effective manifold of hybridized $d$-orbitals centered on the transition metal site in an atomic-like model. Based on either definition, it is clear that \ce{CrI3} does not host strictly defined $dd$-excitations (see Fig.~4 of the main text and Sec.~\ref{ref:atomic_model}).

However, less stringent definitions of $dd$-excitations are sometimes employed in the literature. The term $dd$-excitation is generally less likely to be used for a transition that exhibits interesting coupling to magnetism, but there are exceptions here. The terms exciton and $dd$-excitation are also sometimes associated with narrow and broad linewidths, respectively, but to our knowledge, no attempt has been made to quantify the boundary between these categories. Furthermore, as linewidths depend on material purity and temperature, such distinctions are not necessarily meaningful. The overwhelming majority of papers consider $dd$-excitation local transitions not dispersive quasiparticles. 

Our newly discovered 1.7~eV transitions in \ce{CrI3} are dispersive, exhibit temperature dependence across the magnetic transition, and display narrow linewidths both in general and compared to other transitions in the material. All these properties are more compatible with the use of the term exciton than the typical use of the term $dd$-excitation. Despite the partial ambiguity, we conclude that D1 and D2 are best termed excitons.

\clearpage
\bibliography{refs}